\documentclass[journal]{IEEEtran}
\usepackage{lettrine}

\usepackage{ifpdf}
\ifCLASSINFOpdf
  \usepackage[pdftex]{graphicx}
  \graphicspath{{../pdf/}{../jpeg/}}
  \DeclareGraphicsExtensions{.pdf,.jpeg,.png}
\else
  \usepackage[dvips]{graphicx}
  \graphicspath{{../eps/}}
  \DeclareGraphicsExtensions{.eps}
\fi

\usepackage{cite}
\newcommand{\fig}[4]{ \begin{figure}[#4]
  \centering
   \includegraphics[width=#3\textwidth]{Figures/#1}
   \caption{#2}\label{fig:#1}
  \end{figure}
}

\usepackage{amsthm}
\usepackage{amssymb}
\usepackage{balance}
\usepackage{bm}
\usepackage{eqparbox}
\usepackage{multirow}
\usepackage{mathrsfs} % https://www.ctan.org/pkg/mathrsfs
\usepackage[ruled,vlined]{algorithm2e}
\usepackage{adjustbox}
\usepackage{amsmath}
\usepackage{stmaryrd}

\usepackage{bbold}

\newtheorem{prop}{Proposition}
\newtheorem{lemma}{Lemma}

\usepackage{mathtools, nccmath}

\usepackage{booktabs, siunitx}
\usepackage[svgnames,table]{xcolor}

\usepackage{graphicx}
\usepackage[scaled]{DejaVuSansMono}
\usepackage[T1]{fontenc}
\usepackage{subcaption}

\usepackage{color}
\usepackage{relsize}
\usepackage{mathtools}
\usepackage{soul}

\begin{document}

\title{Optimizing the Coverage Depth in  Coding Schemes for DNA Data Storage}
\author{Elsa Dupraz, \IEEEmembership{Senior Member,~IEEE}, Boyang Zhang, Olivier Boulle, Dominique Lavenier
\thanks{Elsa Dupraz and Boyang Zhang are with IMT Atlantique and Lab-STICC, Brest, France}\thanks{Olivier Boulle and Dominique Lavenier are with University of Rennes, CNRS-IRISA, and Inria, Rennes, France}\thanks{This work was supported by the PEPR MolecularXiv.}}

\maketitle

\begin{abstract}
This paper addresses channel coding for data storage on synthetic DNA. We consider a concatenated scheme in which an inner code corrects the errors introduced by the storage channel, while an outer code compensates for the sampling uncertainty inherent to the reading process. Our goal is to optimize the coding rate allocation between the two codes so as to minimize the average number of reads required to recover the full data. To this end, we resort to Taylor expansions to derive analytical approximations of the average number of reads,  under both uniform sampling, where all molecules are equally likely to be read, and non-uniform sampling, where each molecule has its own sampling probability. These expressions account for a non-zero decoding failure probability for each individual sequence. We then formulate the rate allocation problem and show that the derived theoretical expressions allow us to solve the optimization problem efficiently. Finally, we present optimization results for several combinations of inner codes (LDPC, MGCP, convolutional) and outer codes (Reed–Solomon, Raptor), and show that optimizing the tradeoff between the two codes significantly reduces the read cost.
\end{abstract}
\begin{IEEEkeywords}
DNA data storage, insertion-deletion-substitution channel,  coverage depth, coupon collector problem. 
\end{IEEEkeywords}

\section{Introduction}
Data storage on synthetic DNA molecules has recently emerged as a promising alternative to conventional storage technologies (magnetic tapes, hard
drives, etc.), due to higher density and durability.
In a DNA data storage system, a data item (an image, a text file, etc.) is split into $k$ sequences, and synthesis (the write process) maps each digital sequence onto a DNA molecule. All molecules are stored together in a tube
and replicated by polymerase chain reaction (PCR). Sequencing (the read process) allows one to retrieve the digital sequences. 
%In practice, a data item (an image, a text file, etc.) is split into $k$ sequences, each synthesized as a molecule. 
%All molecules are stored together in a tube
%and replicated by polymerase chain reaction (PCR). 
% 
This storage process is subject to two major sources of uncertainty. First, errors are introduced in the read sequences, due to the synthesis, the PCR, and the sequencing~\cite{gimpel2023digital,hamoum2023channel,heckel2019characterization}. %For instance, the widely used Nanopore sequencer, which is fast and can handle long molecules,
%introduces between $0.5\%$ and $5\%$ of errors, depending on the chemistry and on the basecaller technology.
Second, sampling uncertainty arises because the sequencer reads a random subset of the molecules in the tube, so that some molecules may be read many times while others are not read at all.

The errors introduced in the read sequences are insertions, deletions, and substitutions (IDS). Error-correction codes (ECCs) specifically tailored to IDS errors
have been proposed in the
literature~\cite{davey2001reliable,wang2011symbol,hamoum2023synchronization,shibata2019design,song2022systematic,sima2020optimal,maarouf2022concatenated,khabbaz2025marker,banerjee2024error,vidal2024concatenated,pfister2021polar,chandak2020overcoming}. Each of these codes targets a specific operating point in terms of redundancy and
error-correction capability. For instance, the codes
of~\cite{song2022systematic,sima2020optimal} are low-redundancy codes that only correct a small number of errors, whereas the concatenated scheme based on a convolutional and an LDPC code proposed in~\cite{maarouf2022concatenated} has a very low rate but corrects a large number of errors. The LDPC-based schemes
of~\cite{davey2001reliable,hamoum2023synchronization,shibata2019design,wang2011symbol} and the marker guess-and-check (MGCP) scheme 
of~\cite{khabbaz2025marker} target intermediate rates and channel conditions. 
The ECC is often combined with a consensus algorithm borrowed from bioinformatics, which produces a consensus sequence from $h$ reads of a given molecule and
already removes part of the errors prior to ECC decoding~\cite{park2023reducing,katoh2008recent,de2025bruijn}.

To overcome the sampling uncertainty, an erasure code (ER) can be added as an outer code on top of the ECC, which then acts as an inner code. The outer code generates $n-k$ additional sequences, which are stored as molecules together with the initial $k$ ones, resulting in a pool of $n$ molecules. Reed--Solomon (RS) codes are commonly considered as outer codes~\cite{organick2018random,khabbaz2025marker,bar2024cover,press2020hedges}. Indeed, with RS codes,  collecting any set of $k$ distinct molecules out of $n$ suffices to reconstruct the data. %\st{This significantly reduces the average number of reads, denoted $\mathbb{E}[T]$ and hereafter referred to as the coverage depth.\footnote{\textcolor{red}{Part of the literature defines the coverage depth as the average number of reads per molecule; throughout this paper, it denotes the total number of reads produced by the sequencer.}}}
 %RS decoding,
%however, relies on Gaussian elimination and is computationally expensive. 
%Raptor codes~\cite{shokrollahi2006raptor} are also considered in this context as a less complex yet less efficient alternative~\cite{erlich2017dna}. When including the consensus algorithm, the sequencing problem turns down to collecting $h$ copies of $k$ out of $n$ molecules. 
Raptor codes~\cite{shokrollahi2006raptor,erlich2017dna} and spatially-coupled (SC) LDPC codes~\cite{welter2026end} have also been considered as less complex, yet less efficient, alternatives to RS codes.
%\st{When the consensus algorithm is included, the sequencing problem amounts to collecting $h$ copies of $k$ out of $n$ molecules}.

Both the inner and the outer code impact the write cost and the read cost of the DNA data storage system. The write cost comes from the synthesis process and is essentially set by the number and length of the molecules stored in the tube. 
%which in turn fixes the redundancy budget available to the codes. 
The read cost is driven by the coverage depth, defined as the number of reads per DNA strand that the sequencer must produce before the data can be recovered~\cite{heckel2019characterization}. In what follows, the average number of reads needed to reconstruct the data is denoted by $\mathbb{E}[T]$, and it is proportional to the coverage depth. %\textcolor{blue}{and referred to as the coverage depth}.  
For a fixed synthesis cost, the allocation of redundancy between the two codes involves an inherent tradeoff~\cite{hanna2026reliability,bar2024cover}: reducing the redundancy of the outer code decreases $\mathbb{E}[T]$, but so does reducing the redundancy of the inner code, since incorrectly decoded sequences are discarded. In addition, the number of reads $h$ used by the consensus algorithm also impacts $\mathbb{E}[T]$. 
Therefore, this paper addresses the following question: \emph{given a write cost fixed by the synthesis budget, how should the available redundancy be
allocated so as to minimize the read cost?}

To address this tradeoff, we formulate an optimization problem that consists in minimizing the average number of reads $\mathbb{E}[T]$ with respect to the inner code, outer code, and consensus parameters. 
A similar problem was first formalized as the random access coverage depth problem in~\cite{bar2024cover}, and exact solutions were provided for the error-free case (no error introduced in the reads), and for the noisy singleton case where only one sequence needs to be reconstructed. An instance of the coverage depth problem was also investigated in~\cite{hanna2026reliability}, which considered the noisy case with substitution errors only,  and optimized the sequencing cost to achieve a prescribed reconstruction error probability. Some other works also investigated the tradeoff between the read cost and the write cost for different specific code constructions such as LDPC codes in~\cite{chandak2019improved}, and a combination of SC-LDPC codes and marker-repeat codes in~\cite{welter2026end}. Finally, some other works such as~\cite{gimpel2026comparison} reported the read cost versus write cost of different code construction, for evaluation and comparison rather than optimization. 

Here, we aim to address the general case where the three types of errors can occur, and where all the $k$ sequences need to be reconstructed. We also aim to consider a large class of inner codes and outer codes, so as to optimize different inner/outer code combinations and identify the best setups depending on the operating point (redundancy budget, error rate, etc.). 
The general formulation of the optimization problem is difficult to solve, as it involves the parameters of both codes together with the number $h$ of reads required by the consensus algorithm. Since these variables are discrete, the average number of reads $\mathbb{E}[T]$ must be evaluated over a large search space, and running costly Monte Carlo simulations for each candidate set of parameters is not a viable option.

%\textcolor{blue}{
%\begin{itemize}
%\item %In \cite{hanna2026reliability}: upper/lower bounds on success/failure probability for both uniform and non-uniform case. Optimization in terms of sequencing cost/synthesis cost of the number of reads needed to achieve a prescribed success probability. \textbf{Substitution errors only} 
%\item It's easier to approximate~\cite{tse2010s}
%\item \cite{bar2024cover,bodur2026random} provide $E[T]$ for the case of a singleton (only $1$ sequence needs to be reconstructed). "Random Access coverage depth problem". 
%\item In~\cite{bar2024cover}
 %optimization problem as well: solved for the noiseless case and the singleton case.
% \item  In~\cite{chandak2019improved}, tradeoff between read cost and write cost, and provide an LDPC-based coding scheme that optimally addresses this tradeoff. There are other papers that show results on this tradeoff. 
% \end{itemize}
%}

%\textcolor{red}{Add a bit more of literature in the next paragraph. }
Alternatively, analytical expressions of the average number of reads $\mathbb{E}[T]$ can be obtained from the coupon collector problem, in which one determines the average number of draws required to obtain $h$ copies of $k$ out of $n$ coupons. Exact analytical expressions of $\mathbb{E}[T]$ were provided in~\cite{flajolet1992birthday,brayton1963asymptotic} for both the uniform case, in which all coupons are equally likely, and the non-uniform case, in which the coupons may have different probabilities.
However, these expressions are given in integral form and they are difficult to evaluate numerically. The works~\cite{bar2024cover,bodur2026random} provide simpler analytical expressions of $\mathbb{E}[T]$, but for the singleton case only. 
Several approximations of $\mathbb{E}[T]$ have been proposed to circumvent the computational issue, but none of them is simultaneously tractable and accurate over the range of parameters of interest.
The upper bound of~\cite{bar2024cover}, derived for the uniform case, turns out to be loose in our numerical evaluations.
Recently,~\cite{abraham2026analyzing}, introduced a numerical evaluation method based on dynamic programming for a specific Markov model representing the coupon collector process. 
%The Markov-chain-based numerical method (\textcolor{blue}{Approach based on dynamic programming model for a specific Markov model representing the CCP process}) 
The results of~\cite{abraham2026analyzing} show that this method is accurate, although it is limited to small values of $n$.
Therefore, in this work, we target larger values of $n$ and derive
accurate and tractable approximations of $\mathbb{E}[T]$ for both the uniform and non-uniform cases, by resorting to Taylor expansions.

The main contributions of this paper are as follows:
\begin{enumerate}
\item We derive tractable yet accurate approximations of the average number of reads $\mathbb{E}[T]$ for the general case in which $h$ copies of $k$ out of $n$ coupons are collected. The proposed expressions cover the uniform case, the non-uniform case, and the case in which errors may occur due to incorrect ECC decoding. The accuracy of the proposed expressions is verified by a comparison with Monte Carlo simulations, and we report average relative errors around $0.5\%$ for the least favorable cases (\emph{e.g.}, small $n$, non-uniform case), and less than $0.01\%$ for large values of $n$. 
\item We formulate the problem of minimizing $\mathbb{E}[T]$ with respect to the inner code, outer code, and consensus parameters. We show that the proposed theoretical expressions greatly simplify the resolution of this problem by exhaustive search. Indeed, analytical expressions show that we can separate the contribution of each code into two distinct terms. We first run Monte Carlo simulations to measure the consensus + ECC error probability $\mathbb{P}_e$ independently of the ER code. We then resort to the analytical expression to evaluate $\mathbb{E}[T]$ for the combination of consensus + ECC (only through $\mathbb{P}_e$) and ER.   %: the contribution
%of the sampling process is evaluated analytically, and Monte Carlo simulations are required only to characterize the performance of the ECC alone. %\textcolor{red}%{Pas assez percutant encore?}
\item We report optimization results for several channel error probabilities and redundancy budgets, considering a wide range of ECCs from the literature. These results confirm that the best ECC depends on the operating point, and they highlight the need to jointly optimize the ECC, the ER code, and the consensus parameters in order to reduce the average number of reads. %\textcolor{red}{Can I indicate what we gain, with figures, when using the optimization method?}
\end{enumerate}

The remainder of this paper is organized as follows.
Section~\ref{sec:coding_scheme} presents the coding scheme and the error model considered for DNA data storage. Section~\ref{sec:theoretical_expressions}
derives the theoretical expressions of the average number of reads. Section~\ref{sec:optim_pb} formulates and investigates the optimization problem. Section~\ref{sec:numerical_results} evaluates the accuracy of the proposed expressions and provides solutions to the optimization problem.

%Optimization method based on a theoretical analysis. This methodology is versatile in that it applies to a wide range of erasure codes and error-correction codes. 
\section{Coding scheme for DNA data storage}\label{sec:coding_scheme}

\begin{figure*}[htbp]
\centering
\includegraphics[width=0.85\linewidth]{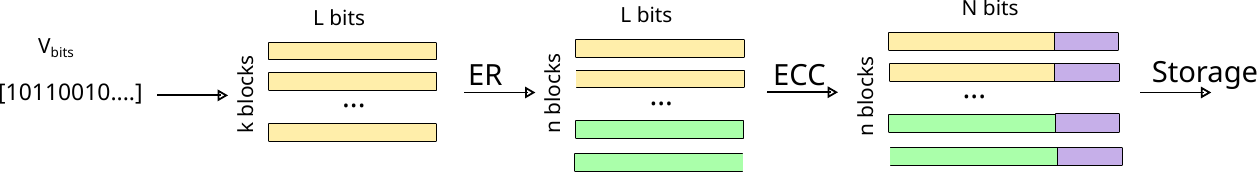}
\caption{Coding scheme for DNA data storage, with one ER and one ECC. The ER adds $(n-k)$ blocks of redundancy (in green), and the ECC adds $(N-L)$ redundancy bits in each block (in purple).  }
\label{fig:coding_scheme}
\end{figure*}

Throughout the paper, we consider the storage of $V_{b}$ bits of information. We assume that the synthesis produces $n$ DNA molecules, each of length $N$ bits, or equivalently $\lceil N/2 \rceil$ quaternary symbols. In addition, we consider $V_b < n\times N$, to leave some space for the redundancy of the codes. We also assume that the values of $n$ and $N$ are fixed by the synthesis process. After synthesis, the $n$ molecules are stored together in a tube, and replicated using a PCR. The sequencing produces reads of the molecules, and from these reads, one aims to reconstruct exactly the $V_{b}$ bits of information. %In this section, we describe the considered DNA storage channel model, as well as the coding scheme with inner and outer codes. 

\subsection{DNA storage channel model}
%To describe uncertainties introduced by the storage process, we consider one sampling model for the selection of the molecules during sequencing, and one error model to represent errors introduced by the storage process.  
%In DNA data storage, synthesis (the write process) transforms a digital sequence into a DNA molecule. All DNA molecules are stored together in a tube, and they are possibly amplified with a PCR. Sequencing (the read process) consists of reading the DNA molecules into digital sequences. To describe this process, we consider two statistical models: one sampling model for the selection of the molecules during sequencing, and one error model to represent errors introduced by the process. 
%\subsubsection{Sampling model}
%As sampling model, 

We consider that each sequencing read consists of sampling with replacement one of the $n$ DNA molecules. We assume that molecule $i \in \llbracket 1,n\rrbracket$ is selected with probability $p_i$, where $\sum_{i=1}^n p_i = 1$, and we use $\mathbf{p}=(p_1,\cdots,p_n)$ to denote the sampling distribution. In this paper, we will consider both the non-uniform case where each $p_i$ can differ, and the uniform case where $p_i = 1/n$ for all $i$. In practice, the values of $p_i$ are not uniform~\cite{gimpel2023digital,heckel2019characterization} and depend on the synthesis, PCR, and sequencing. %Here, for simplicity, the distribution $\{p_i\}_{i \in \llbracket 1,N\rrbracket}$ is assumed known from the beginning. 
We further consider that the three types of errors (insertion, deletions, substitutions) can occur in the read sequences, and we assume only transient (\emph{i.e.}, non-systematic) errors. Under this assumption, the proposed optimization method encompasses both i.i.d. and non-i.i.d. error models~\cite{hamoum2023channel,maarouf2023achievable,mcbain2022finite} with any error statistics, and the error model will have an effect only through the performance of the considered ECC. In our theoretical analysis, we will consider both the error-free case where uncertainty only comes from the sampling, and the noisy case where IDS errors are introduced in the reads.  

%Current synthesis technologies introduce negligible error rates compared to sequencing. In contrast, sequencing platforms such as Nanopore and Illumina introduce IDS errors. 
%Many existing works either assume that the DNA storage process introduces a maximum number of errors per sequence~\textcolor{red}{[REF]}, or consider a statistical model with independent and identically distributed (i.i.d.) errors~\textcolor{red}{[REF]}. 
%As a more realistic alternative, we adopt the channel model introduced in~\cite{hamoum2023channel} and considered in~\cite{maarouf2023achievable}, which targets Nanopore sequencing and captures both the statistics and memory of sequencing errors through a hidden Markov model. This model was trained on multiple datasets, yielding different error profiles with error rates ranging from $0.9\%$ to $10\%$. Note that in this work, the error model does not impact the theoretical developments, and we consider Nanopore sequencing as a relevant use-case to provide numerical optimization results. The optimization method can be implemented with any other error model. 

\subsection{Coding scheme}

\begin{figure*}[htbp]
\centering
\includegraphics[width=0.8\linewidth]{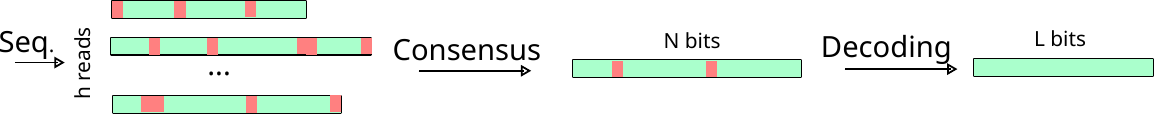}
\caption{Reconstruction scheme for DNA data storage, for one sequence, applying a consensus over $h$ reads of the sequence, followed by ECC decoding.  Error positions are shown in red.   }
\label{fig:rec_scheme}
\end{figure*}

To reliably reconstruct information from the noisy and randomly sampled reads, a two-layer coding scheme is employed, as depicted in Figure~\ref{fig:coding_scheme}. 
The $V_{\text{b}}$ data bits are divided into $k$ sequences of $L$ bits each, so that $V_{\text{b}} = k \times L$. These sequences are first encoded using an ER code as outer code. This code produces $k_r$ additional redundant sequences of length $L$, resulting in $k+k_r \leq n$ sequences of the same length~$L$. Outer codes are often RS codes~\cite{organick2018random,khabbaz2025marker,bar2024cover,press2020hedges}, or Raptor codes~\cite{shokrollahi2006raptor,schwarz2024data}. 
Each sequence is then further protected by an ECC used as an inner code, which encodes each sequence of length $L$ into a codeword of length $L + L_r \leq N$, where $L_r$ is the redundancy added by the ECC. In the case where $k+k_r < n$, or $L+L_r < N$, dummy bits are added to reach $n$ sequences of $N$ bits each. The resulting $n$ codewords of length $N$ are synthesized to produce $n$ DNA strands.
ECC solutions developed for DNA data storage include convolutional codes~\cite{maarouf2022concatenated}, MGCP codes~\cite{khabbaz2025marker}, LDPC codes~\cite{davey2001reliable,hamoum2023synchronization,shibata2019design,wang2011symbol}, and Polar codes~\cite{pfister2021polar}.

%\subsection{Data reconstruction}
After sequencing, the stored information is recovered through a three-stage process illustrated in Figure~\ref{fig:rec_scheme}. First, a consensus algorithm produces a consensus sequence from $h$ reads of the same DNA strand, without exploiting the ECC. Existing consensus methods include majority voting based on multiple alignment~\cite{katoh2008recent}, as well as CCSA~\cite{de2025bruijn} and OLC algorithms. After consensus, the ECC decoder corrects residual errors. Once a sufficient number of sequences have been reconstructed without error, the ER decoder retrieves the full information. For example, for RS codes, collecting any set of $k$ correctly reconstructed distinct sequences is sufficient to retrieve the $n$ coded sequences. In what follows, we implicitly assume that after ECC decoding, errors are detected using a cyclic-redundancy check (CRC) in all sequences. However, we do not consider the CRC in our optimization, as it impacts all considered ECC solutions in the same way. 

%\textcolor{red}{$h$ fixed is suboptimal... this should be considered as a simplifying assumption. }

\subsection{Expected number of reads}
Denote by $T$ the number of reads needed to reconstruct the $V_b$ bits of information without error. The quantity $T$ is proportional to the coverage depth defined as $T/n$.  Since $T$ is a random variable, we are more interested in its expectation $\mathbb{E}[T]$. 
The value of $\mathbb{E}[T]$ depends on the parameters $n,k,h$, the sampling distribution $\mathbf{p}$, and the channel error model.
Therefore, in the non-uniform case, we denote $\mathbb {E}[\widetilde{T}_{\mathbf{p}}(n,k,h)]$ and $\mathbb {E}[T_{\mathbf{p}}(n,k,h)]$ the average number of reads in the error-free and noisy cases, respectively. In the uniform case, we simplify the notation to $\mathbb{E}[\widetilde{T}(n,k,h)]$ and $\mathbb {E}[T(n,k,h)]$, respectively.  
Both exact and approximate expressions of these quantities are provided in the literature on the coupon collector problem.  %The uniform assumption constitutes a necessary first step to develop the proposed optimization framework and to compare different coding solutions. Recent works have investigated $\mathbb{E}[T]$ under non-uniform distributions~\cite{abraham2026analyzing}, but the numerical evaluation of the resulting expressions remains intractable for large~$n$. 

First assume error-free sequencing, no consensus, and consider a uniform read distribution. Under these assumptions and  when no ER is implemented, all $k$ original sequences must be sampled at least once, requiring on average~\cite{newman1960double,flajolet1992birthday}
\begin{equation}
\mathbb{E}[\widetilde{T}(k,k,1)] = kH_k \approx k\log k~ \text{ reads},
\end{equation}
where $H_k$ denotes the $k$-th harmonic number. When a RS code is used as ER, one only needs to collect $k$ unique sequences among the $n$ encoded ones. This reduces the expected number of reads to~\cite{flajolet1992birthday}
\begin{equation}\label{eq:approx_h1}
\mathbb{E}[\widetilde{T}(n,k,1)] = n(H_n - H_{n-k}) \approx n \log \frac{n}{n-k}.
\end{equation}
Since $n(H_n - H_{n-k}) \leq kH_k$, the ER reduces the expected number of reads needed to reconstruct the data.  

Next, still in the error-free case, when considering $h$ reads and a non-uniform distribution, an exact analytical expression is given in~\cite{flajolet1992birthday}:
\begin{equation}\label{eq:ET_nonunif_integral}
\begin{split}
\mathbb{E}[\widetilde{T}_{\mathbf{p}}(n,k,h)] = \sum_{q=0}^{k-1} \int_0^\infty [u^q] \prod_{i=1}^{n} \Big( e_{h-1}(p_i v) \\
+ \, u \cdot \left( e^{p_i v} - e_{h-1}(p_i v) \right) \Big) e^{-v} \, dv,
\end{split}
\end{equation}
where 
%\begin{equation}
$e_h(x) = \sum_{i=0}^{h} \frac{x^i}{i!},$
%\end{equation}
and $[u^q]Q(u)$ is the coefficient of $u^q$ in the polynomial $Q(u)$. 
The expression of $\mathbb{E}[\widetilde{T}(n,k,h)]$  for the uniform case is obtained by substituting $p_i = 1/n$ in~\eqref{eq:ET_nonunif_integral}, which does not bring any further simplification to the integral. 

The integral form in~\eqref{eq:ET_nonunif_integral} is numerically intractable for large~$n$~\cite{boneh1997coupon}. A more efficient evaluation method, accounting for the underlying Markov model of the process, is proposed in~\cite{abraham2026analyzing}, but still the approach does not scale well as $n$ grows. %, and in~\cite{abraham2026analyzing}, numerical evaluation was limited to $n=160$.
Furthermore, in the uniform case,~\cite[Theorem~5]{bar2024cover} provides the following upper bound: 
\begin{equation}\label{eq:upper_bound}
\mathbb{E}[\widetilde{T}(n,k,h)]  \leq n\left(\log \frac{n}{n-k} +  h \log \log n + 2\log(h+1)\right). 
\end{equation}
However, our numerical results show that this bound is loose. 
Alternatively, in what follows, we provide accurate yet tractable approximated expressions of $\mathbb{E}[\widetilde{T}_{\mathbf{p}}(n,k,h)]$ in the uniform and non-uniform cases. Besides, the ECC error probability also affects $\mathbb{E}[T]$. In particular, since the ER decoder runs only from perfectly reconstructed sequences, any sequence that the ECC fails to decode correctly is discarded, thereby increasing the expected number of reads. Therefore, in what follows, we also provide expressions of $\mathbb{E}[T_{\mathbf{p}}(n,k,h)]$ in the noisy case, taking into account the ECC error probability $\mathbb{P}_e$. 

%Finally, it is worth mentioning that other special cases are addressed in the literature. 
%For example,~\cite{newman1960double} provides an approximate expression of $\mathbb{E}[\widetilde{T}(n,n,h)]$ for the uniform case with $k=n$. And~\cite{cao2026modeling} provides an approximate expression of  $\mathbb{E}_{\mathbf{p}}[\widetilde{T}(n,n,1)]$ for the non-uniform case with $h=1$.  %And approximate expression for the non-uniform case with $h=1$, and~\cite{bar2024cover} provides an exact expression for this case. 
%$E[T] = n\log(n) + n(h-1) \log(\log(n) $). In~\cite{cao2026modeling}, approximate expression for the non-uniform case with $h=1$, and~\cite{bar2024cover} provides an exact expression for this case.  

%Alternatively, the following propositions provide approximations that are both accurate and computationally simple. We consider both the uniform and non-uniform cases. 

%In the remaining of the paper, we first provide analytical expressions for the expected number of reads $\mathbb{E}[T]$ in setups of interest not addressed in the literature. We then address the optimization problem that consists of minimizing $\mathbb{E}[T]$ with respect to the code parameters. 

\section{Theoretical expressions of the average number of reads}\label{sec:theoretical_expressions}
%In this section, we derive accurate approximations of $\mathbb{E}[T]$. We restrict attention to RS codes as ER, for which any $m$ error-free sequences among the $n$ encoded ones suffice to recover the data.

\subsection{Error-free case}\label{sec:th_errorfree}
We first consider the error-free case and assume that all sequences are correctly decoded after consensus and ECC decoding. Our goal is to derive an analytical expression for $\mathbb{E}[\widetilde{T}_{\mathbf{p}}(n,k,h)]$, the expected number of reads required to collect $h$ copies of $k$ distinct sequences out of $n$. 
Throughout this section, $h$ is fixed, while $k=k(n)$ and $\mathbf{p}=\mathbf{p}^{(n)}$ depend on $n$. Asymptotic notations refer to $n\to\infty$, and the constants depend only on $h$, $\varepsilon$, $c_1$, and $c_2$ defined below.
%Throughout this section, $h$ is fixed, $k=k(n)$ depends on $n$, and asymptotic notations refer to $n\to\infty$. 
 
\begin{prop}[uniform, error-free  case]\label{prop:average_T}
Assume that each molecule is sampled with equal probability $p_i = 1/n$ for all $i \in \llbracket 1,n\rrbracket$, and that there exist $\varepsilon>0$ such that for all $n$, $k/n\in[\varepsilon,1-\varepsilon]$. The expected number of reads required to collect $h$ copies of $k$ distinct sequences out of $n$ satisfies
\begin{equation}\label{eq:Et_uniform}
\mathbb{E}[\widetilde{T}(n,k,h)]  = \, \mathbb{F}^{-1}\!\left(\frac{k}{n+1}\right) + O(1),
\end{equation}
where $\mathbb{F}$ is the cumulative distribution function of the Gamma distribution $\Gamma(h,n)$.  
\end{prop}
\begin{proof}
See Appendix~\ref{app:prop1}. 
\end{proof}
%Note that~\cite[Section~5.5]{arnold2008first} also provides a Taylor expansion for the variance of $T_{k:n}$. 

%Let $\mathbf{p}$ denote the sampling distribution. Assume that there exists $\varepsilon>0$ such that for all $n$, $k/n\in[\varepsilon,1-\varepsilon]$, and that there exists $c_1,c_2>0$ such that $c_1/n\le p_j\le c_2/n$ is verified for all $j\in \llbracket 1,n\rrbracket$. The expected number of reads required to collect $h$ copies of $k$ distinct sequences out of $n$ satisfies
\begin{prop}[non-uniform, error-free case]\label{prop:Et_nonunif}
Let $\mathbf{p}$ denote the sampling distribution. Assume that there exists $\varepsilon>0$ such that, for all $n$, $k/n\in[\varepsilon,1-\varepsilon]$, and that there exist $c_1,c_2>0$ such that, for all $n$ and all $j\in \llbracket 1,n\rrbracket$, $c_1/n\le p_j\le c_2/n$. The expected number of reads required to collect $h$ copies of $k$ distinct sequences out of $n$ satisfies
\begin{align}\label{eq:Et_nonuniform}
\mathbb{E}[\widetilde{T}_{\mathbf{p}}(n,k,h)]   = \mathbb{F}_{\mathbf{p}}^{-1}\!\left(\frac{k}{n+1}\right) + O(\sqrt{n\log n}),
\end{align}
where $\mathbb{F}_{\mathbf{p}}$ is a cumulative distribution function defined by
%\begin{equation}\label{eq:distrib_nonunif}
%\mathbb{F}_{\mathbf{p}}(x) = \sum_{j=1}^n \frac{1}{n}\, \mathbb{F}_{(h,\,1/p_j)}(x),
%\end{equation}
\begin{equation}\label{eq:distrib_nonunif}
\mathbb{F}_{\mathbf{p}}(x) = \frac{1}{n}\sum_{j=1}^n \, \mathbb{F}_{j}(x),
\end{equation}
and $\mathbb{F}_{j}$ denotes the cumulative distribution function of the Gamma distribution $\Gamma(h,1/p_j)$. 
%\textcolor{red}{$1/n$ instead of $p_i$?}
\end{prop}
\begin{proof}
See Appendix~\ref{app:prop2}. 
\end{proof}
%Note that, unlike the uniform case~\eqref{eq:Et_uniform}, the expression~\eqref{eq:Et_nonuniform} does not include the multiplicative factor~$n$, since the non-equal sampling probabilities are already absorbed into the scale parameters of the Gamma distributions in~\eqref{eq:distrib_nonunif}. 

Interestingly, it is shown in~\cite{bar2024cover} that the uniform case minimizes the average number of reads, \emph{i.e.}, $\mathbb{E}[\widetilde{T}(n,k,h)]  \leq \mathbb{E}[\widetilde{T}_{\mathbf{p}}(n,k,h)]$. This shows that both $\mathbb{E}[\widetilde{T}(n,k,h)] $ and $\mathbb{E}[\widetilde{T}_{\mathbf{p}}(n,k,h)]$ grow as $n$ (or faster) according to~\eqref{eq:approx_h1} and given that $\mathbb{E}[\widetilde{T}_{\mathbf{p}}(n,k,h)]$ is increasing with $h$. We also note that $\mathbb{E}[\widetilde{T}_{\mathbf{p}}(n,k,h)] $ in Proposition~\ref{prop:Et_nonunif} has a larger residue, in $O(\sqrt{n\log n})$, than the one of  $\mathbb{E}[\widetilde{T}(n,k,h)]$ in Proposition~\ref{prop:average_T}. This comes from the different proof techniques considered in both Propositions. 
The accuracy of the expressions provided in Propositions~\ref{prop:average_T} and~\ref{prop:Et_nonunif} will be validated against Monte Carlo simulations in Section~\ref{sec:numerical_results}. Note that the condition $k/n\in[\varepsilon,1-\varepsilon]$ is due to the Taylor expansion, and reflects the fact that the approximation is inaccurate for values of $k$ close to $0$ and close to $n$. 

\subsection{Noisy case}\label{subsec:Et_noisy}
We now consider the noisy case in which each sequence, after consensus and ECC decoding, has a block error probability $\mathbb{P}_e$. %We want to express $\mathbb{E}[T(n,k,h)]$. 
In this case, a key design choice is how to handle incorrectly decoded sequences. %and re-attempting consensus with $h$ fresh reads (Strategy~3).
Here, we consider that if the ECC decoder fails for sequence $i \in \llbracket 1,n\rrbracket$, then the $h$ reads used for the failed consensus are discarded. If $h$ new reads of sequence~$i$ are subsequently collected, a new consensus attempt is made.

\begin{prop}[uniform, noisy case]\label{prop:Et_strat3}
Assume that each molecule is sampled with equal probability $p_i = 1/n$ for all $i \in \llbracket 1,n\rrbracket$, and that there exists $\varepsilon>0$ such that for all $n$, $k/n\in[\varepsilon,1-\varepsilon]$. Upon failure of ECC decoding with probability $\mathbb{P}_e<1$, assume that the $h$ reads of the sequence are discarded. Then, the expected number of reads required to collect $k$ correctly decoded sequences out of $n$ satisfies
\begin{equation}\label{eq:Et_strat3}
\mathbb{E}[T(n,k,h)] = \,\mathbb{G}^{-1}\!\left(\frac{k}{n+1}\right) + O(1),
\end{equation}
where $\mathbb{G}$ is the cumulative distribution function defined by
\begin{equation}\label{eq:G_cdf}
\mathbb{G}(x) = (1-\mathbb{P}_e) \sum_{r \geq 1} \mathbb{P}_e^{r-1}\,\mathbb{F}_{r}(x),
\end{equation}
and $\mathbb{F}_{r}$ is the cumulative distribution function of the Gamma distribution $\Gamma(rh,n)$. 
\end{prop}
\begin{proof} See Appendix~\ref{app:prop3}. 
\end{proof} 

\begin{prop}[non-uniform, noisy case]\label{prop:Et_strat4}
Let $\mathbf{p}$ denote the sampling distribution. Assume that there exists $\varepsilon>0$ such that, for all $n$, $k/n\in[\varepsilon,1-\varepsilon]$, and that there exist $c_1,c_2>0$ such that, for all $n$ and all $j\in \llbracket 1,n\rrbracket$, $c_1/n\le p_j\le c_2/n$.  Upon failure of ECC decoding with probability $\mathbb{P}_e<1$, assume that the $h$ reads of the sequence are discarded. Then, the expected number of reads required to collect $k$ correctly decoded sequences out of $n$ satisfies
\begin{equation}\label{eq:Et_strat4}
\mathbb{E}[T_{\mathbf{p}}(n,k,h)] = \,\mathbb{G}^{-1}_{\mathbf{p}}\!\left(\frac{k}{n+1}\right) + O(\sqrt{n\log n}),
\end{equation}
where $\mathbb{G}_{\mathbf{p}}$ is the cumulative distribution function defined by
\begin{equation}\label{eq:G_cdf_Pe}
\mathbb{G}_{\mathbf{p}}(x) = (1-\mathbb{P}_e) \sum_{r \geq 1} \mathbb{P}_e^{r-1}\sum_{j=1}^n \frac{1}{n}\, \mathbb{F}_{(rh,\,1/p_j)}(x),
\end{equation}
%\begin{equation}\label{eq:G_cdf}
%\mathbb{G}_{\mathbf{p}}(x) = (1-\mathbb{P}_e) \sum_{r \geq 1} \mathbb{P}_e^{r-1}\,\mathbb{F}_{\mathbf{p},r}(x),
%\end{equation}
and $\mathbb{F}_{(rh,\,1/p_j)}$ denotes the cumulative distribution function of the Gamma distribution $\Gamma(rh,1/p_j)$.  
\end{prop}
\begin{proof} The proof is immediate by combining the proofs of Propositions~\ref{prop:Et_nonunif} and~\ref{prop:Et_strat3}. 
\end{proof}

\subsection{Raptor codes}
For RS codes, collecting exactly $k$ unique sequences  out of $n$ suffices to recover the full information, and it is shown in~\cite{bar2024cover} that as maximum Distance Separable (MDS) codes, they minimize the average number of reads over all possible outer codes. 
However, RS decoding is complex. Alternatively, one can consider suboptimal codes such as Raptor codes~\cite{shokrollahi2006raptor,erlich2017dna}, which have a decoding complexity linear in $n$. 
For Raptor codes, the required number of sequences needed to recover the full information exceeds $k$ and is a random variable. The probability distribution of this random variable depends both on the Raptor code construction and on the specific set of recovered sequences.
Therefore, we let $\mathbb{E}[T_R(n,k,h)]$ denote the average number of reads required to recover the full information when considering Raptor codes. 
Suppose that a probability distribution $\pi_m, m \in \llbracket k,n\rrbracket$, is available, where $\pi_m$ is the probability that exactly $m$ distinct sequences are required for Raptor reconstruction. The expected number of reads can then be expressed as
\begin{equation}\label{eq:T_raptor}
\mathbb{E}[T_R(n,h)] = \sum_{m=k}^n \pi_m \mathbb{E}[T_{\mathbf{p}}(n,m,h)] .
\end{equation}
%where $\mathbb{E}[T(n,m,h)]$ is taken from the previous propositions. %and
%the probability distribution $ \mathbb{P}(k)$ can be estimated via Monte-Carlo simulations.  
%The distribution $\mathbb{P}(k)$ can be evaluated with Monte-Carlo simulations.
% 
%In practice, it is often more convenient to estimate the decoding failure probability $\mathbb{Q}_e(m)$ of the Raptor code for all $m\in \llbracket k,n\rrbracket$, where $\mathbb{Q}_e(m)$ is defined as the probability of decoding failure when exactly $m$ distinct sequences are available to the Raptor decoder. The distribution $\mathbb{P}(m)$ can then be expressed as
%\begin{equation}
%\mathbb{P}(m) = (1-\mathbb{Q}_e(m)) \prod_{i=k}^{m-1} \mathbb{Q}_e(i),
%\end{equation}
%and substituted into~\eqref{eq:T_raptor}. 
In our numerical results, the distribution $\pi_m$ will be estimated from Monte Carlo simulations, and $\mathbb{E}[T_R(n,h)]$ will be evaluated using the expressions of $\mathbb{E}[T_{\mathbf{p}}(n,k,h)]$ from the previous propositions.

\section{Optimization problem}\label{sec:optim_pb}
According to the previous section, the inner code, the outer code, and the consensus algorithm jointly determine the average number of reads $\mathbb{E}[T_{\mathbf{p}}(n,k,h)]$. Therefore, we now investigate the problem of optimizing these codes and their parameters so as to minimize $\mathbb{E}[T_{\mathbf{p}}(n,k,h)]$. 

\subsection{Optimization problem formulation}\label{sec:tradeoffs}

We consider the number of DNA strands~$n$ and the strand length~$N$ (in bits) as fixed parameters, constrained by the synthesis process. %In practice, increasing either $n$ or $N$ incurs higher synthesis costs. 
Under these constraints, the degrees of freedom are: (i) the code parameters $k$ and $L$ such that $k \times L \geq V_{\text{b}}$, (ii) the redundancies of the outer and inner codes $k_r$ and $L_r$ such that $k+k_r \leq n$ and $L+L_r \leq N$, (iii) the parameter $h$ of the consensus algorithm. 
We propose to minimize the average number of reads $\mathbb{E}[T_{\mathbf{p}}(n,k,h)]$ required to perfectly reconstruct the stored data. This yields the following optimization problem:
\begin{equation}\label{eq:opt_problem}
        \displaystyle  \min_{h,k,L,k_r,L_r} ~ 
       \quad  \mathbb{E}[T_{\mathbf{p}}(n,k,h)] \quad \text{s.t.}  \quad 
        \begin{cases}
            k \times L \geq V_{\text{b}}, \\
            k + k_r \leq n, \\
            L + L_r \leq N. 
        \end{cases}
\end{equation}
This formulation accounts for the outer code through the parameters~$k$ and $k_r$, the inner code through the parameters~$L$ and $L_r$, and the consensus depth~$h$. In practice, we simplify this optimization problem by setting $k+k_r = n$, $L+L_r=N$, and optimizing with respect to $k$, $L$, and $h$.   

Solving the optimization problem~\eqref{eq:opt_problem} allows us to investigate three key design questions:
\begin{enumerate}
    \item \textit{Choice of the coding and consensus components.} Different ER codes, ECCs, and consensus algorithms achieve different performance depending on the operating regime defined by $n$, $N$, and $V_{\text{b}}$. We seek to choose the ones that minimize $\mathbb{E}[T_{\mathbf{p}}(n,k,h)]$ for the considered regime. 
    \item \textit{Tradeoff between inner and outer code redundancy.} For a fixed product $k \times L = V_{\text{b}}$, decreasing $k$ allocates more redundancy to the outer code, whereas decreasing $L$ strengthens the error-correction capability of the inner code. We would like to find the pair $(k,L)$ that minimizes $\mathbb{E}[T_{\mathbf{p}}(n,k,h)]$. 
    \item \textit{Tradeoff between consensus depth and ECC capability.} Reducing~$h$ increases the error rate at the output of the consensus algorithm, thereby requiring a more powerful ECC. We would like to find the pair $(h,L)$ that minimizes $\mathbb{E}[T_{\mathbf{p}}(n,k,h)]$. 
\end{enumerate}

%\textcolor{blue}{Note that considering the partial decoding strategy presented in Section xx does not change the optimization problem. It only updates the redundancy $L_r$ as the sum of the redundancy of the inner code and of the term $Jl_s$.  } 

\subsection{Solving the optimization problem}
A key difficulty in addressing the optimization problem~\eqref{eq:opt_problem} lies in the numerical evaluation of the quantity $\mathbb{E}[T_{\mathbf{p}}(n,k,h)]$. Since this quantity depends on the inner code, the outer code, and the consensus algorithm jointly, a brute-force approach would require running Monte Carlo simulations for every code configuration and every combination of parameters $(h,k,L)$. Such an approach is computationally prohibitive given the large search space. %In the following sections, we show that the evaluation of $\mathbb{E}[T]$ can be simplified by resorting to analytical expressions and by separating the contributions of the ECC and the ER. 
Alternatively, the expressions derived in Section~\ref{sec:theoretical_expressions} greatly simplify the evaluation of $\mathbb{E}[T_{\mathbf{p}}(n,k,h)]$ in the optimization problem~\eqref{eq:opt_problem}. Especially, the results of Section~\ref{subsec:Et_noisy} express $\mathbb{E}[T_{\mathbf{p}}(n,k,h)]$ in a form that separates the contribution of the outer code, through $\mathbb{E}[\widetilde{T}_{\mathbf{p}}(n,k,h)]$, from that of the inner code, through $\mathbb{P}_e$. Therefore, we proceed as follows to evaluate the terms $\mathbb{E}[T_{\mathbf{p}}(n,k,h)]$: % for fixed parameters $n,k,h$:
\begin{enumerate}
    \item For the considered ECC  and consensus algorithm, we run Monte Carlo simulations to evaluate the error probabilities $\mathbb{P}_e(L,h)$ for all considered pairs $(L,h)$. 
    \item For the considered ER code, we use the theoretical expressions provided in Section~\ref{sec:th_errorfree} to evaluate $\mathbb{E}[\widetilde{T}_{\mathbf{p}}(n,k,h)]$ in the error-free case, for all considered pairs $(k,h)$. 
    \item We use the results of Section~\ref{subsec:Et_noisy} to evaluate $\mathbb{E}[T_{\mathbf{p}}(n,k,h)]$ from $\mathbb{P}_e(L,h)$ and $E[\widetilde{T}_{\mathbf{p}}(n,k,h)]$, by taking $L$ from the condition $k \times L \geq V_b$. 
\end{enumerate}
Once this is done, we retain the set of parameters $(h,k,L)$ that minimizes $\mathbb{E}[T_{\mathbf{p}}(n,k,h)]$ over all considered configurations. 

We note that, since all quantities involved in the optimization process are discrete, the search is performed exhaustively over the parameter space. A less complex approach such as an alternating optimization over individual parameters could further reduce the computation time, but at the risk of converging to local minima. Given that the analytical formulation already makes the exhaustive search tractable, we opt for exhaustive search to ensure that no favorable configuration is overlooked.

%\subsection{Complexity analysis}
%We use $\overline{h}$, $\overline{k}$, %and $\overline{L}$ to denote the number of candidate values for $h$, $k$, and $L$, respectively. The brute-foce strategy that runs Monte-Carlo simulations of the full system (inner code, outer code, and consensus) for every parameter combination requires a computation time 
%\begin{equation}
%\tau_{\text{BF}} = \overline{h}\,\overline{k}\,\overline{L}\,\tau_s,
%\end{equation}
%where $\tau_s$ is the time needed to simulate one configuration.
%In our case, $\mathbb{E}[\widetilde{T}_{\mathbf{p}}(n,k,h)]$ is evaluated efficiently from the expressions provided in Section~\ref{sec:theoretical_expressions}, at a cost $\tau_t \ll \tau_s$ per configuration. The ECC contribution only requires simulating the ECC decoder (without the ER), at a cost $\tau_s' \leq \tau_s$. The total optimization time therefore reduces to 
%\begin{equation}
%\tau_{\text{A}} = \overline{h}\,\overline{k}\,\tau_t + \overline{h}\,\overline{L}\,\tau_s'.
%\end{equation}
%Since $\tau_t$ is negligible compared to $\tau_s'$, the effective cost is dominated by $\overline{h}\,\overline{L}\,\tau_s'$, and hence $\tau_{\text{A}} \ll \tau_{\text{BF}}$. 

\section{Numerical Results}\label{sec:numerical_results}
\subsection{Accuracy of the theoretical analysis}
\fig{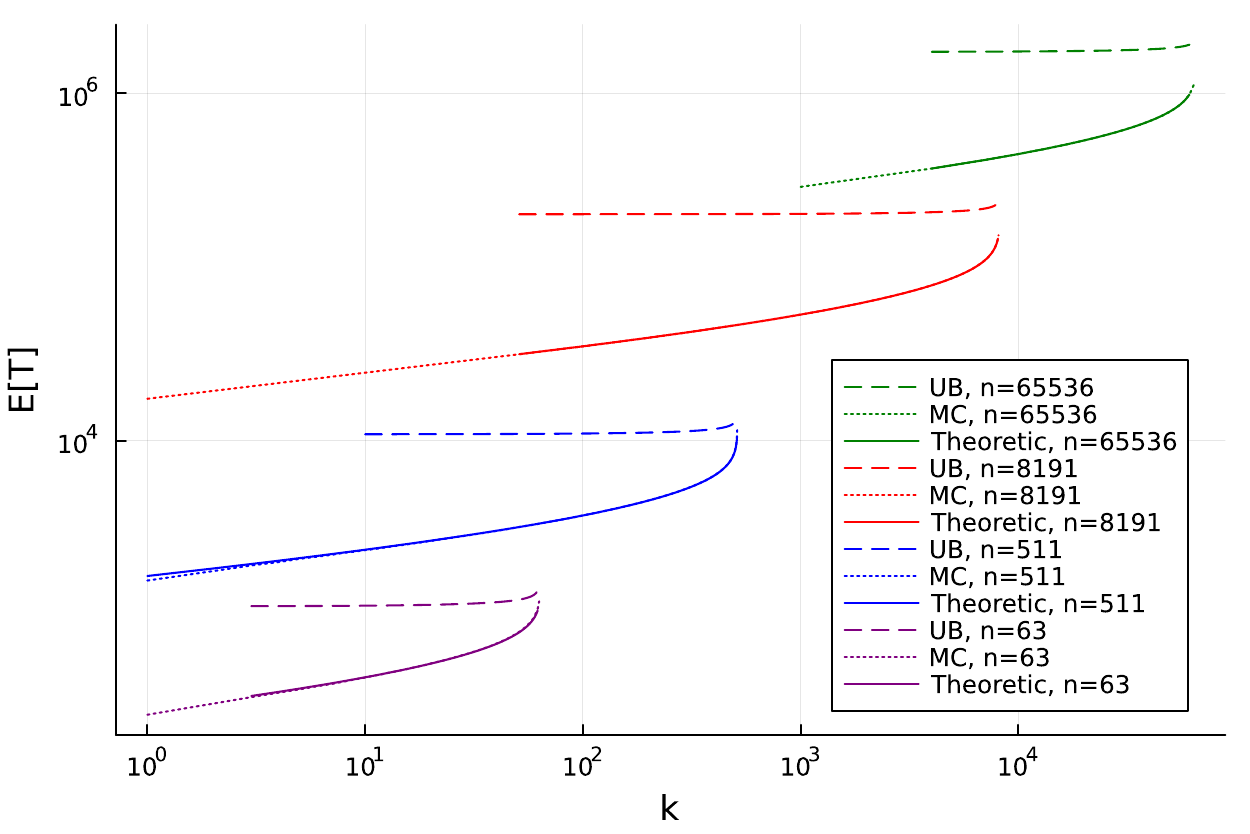}{Comparison of the proposed expressions with Monte Carlo simulations for $n\in \{63,511,8191,65536\}$ and $h=10$. For each value of $n$, we provide the upper bound (UB), the expression from Proposition~\ref{prop:average_T} (Approx), and the Monte Carlo simulations (MC). For each considered case, the theoretical expression coincides with with the Monte Carlo simulations.   }{0.45}{t}

%\fig{RS_partial}{Comparison of the proposed expression with Monte-Carlo simulations for partial RS, with $n=512$, $k=256$, and several values for the number $J$ of sub-blocks.}{0.45}{t}

\fig{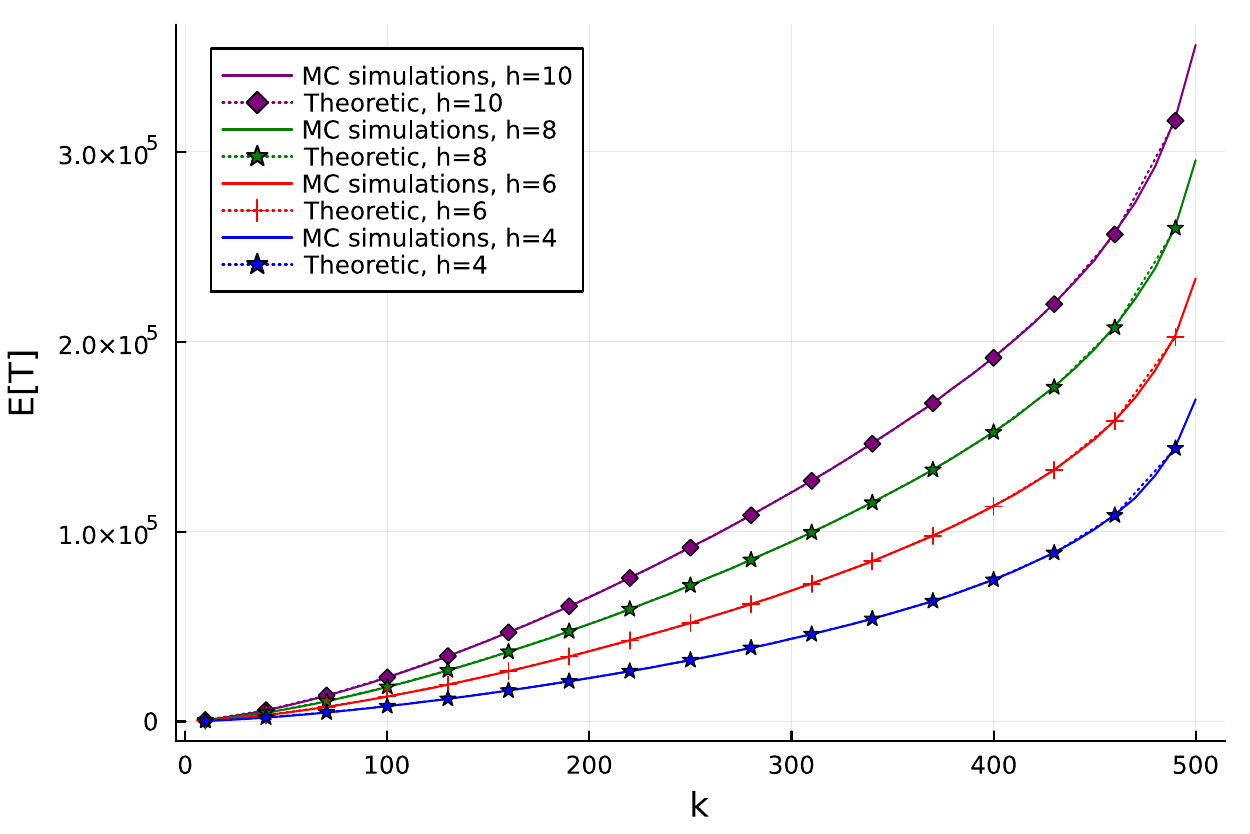}{Comparison of the proposed expression of Proposition~\ref{prop:Et_nonunif}, with Monte Carlo simulations, for a non-uniform Zipf probability distribution over the sequences, in the error-free case,  for $n=511$ and different values of $k$ and $h$. }{0.45}{t}

\fig{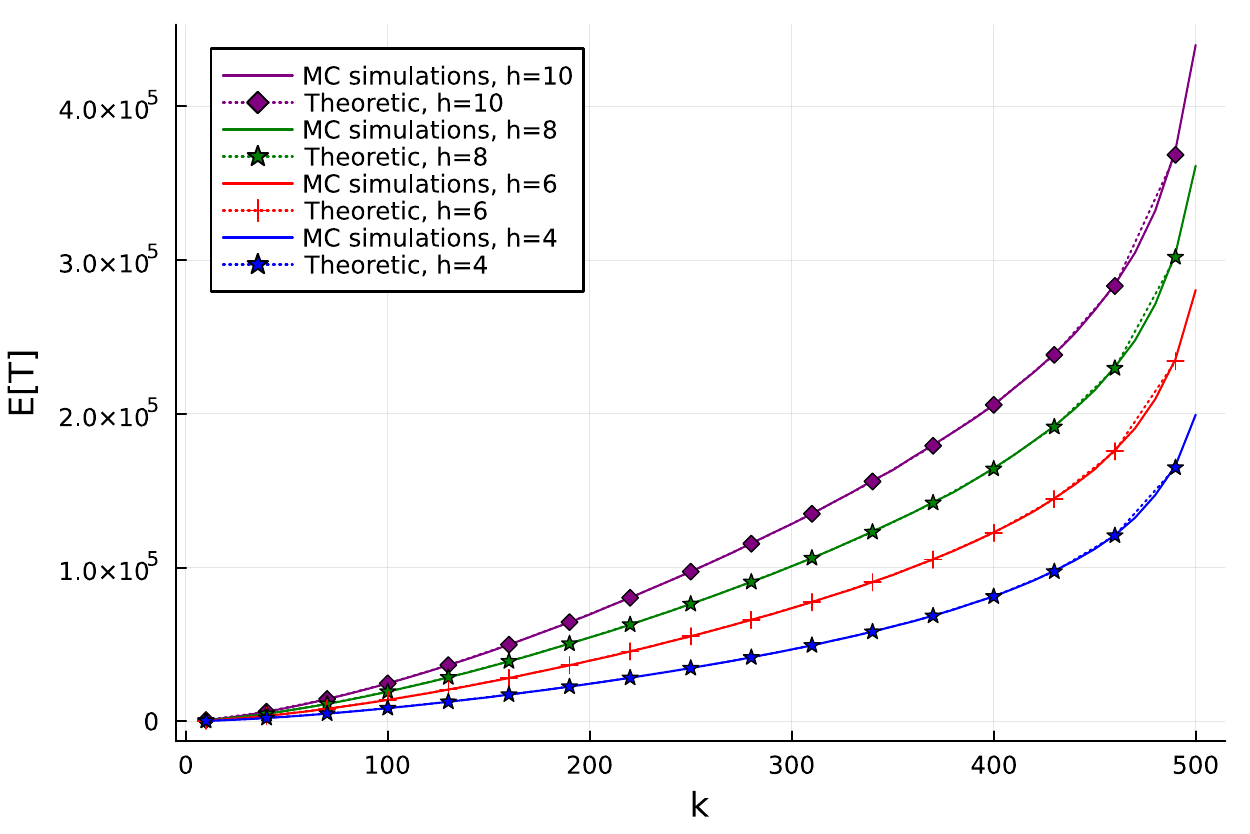}{Comparison of the proposed expression of~\eqref{eq:Et_strat3}, with Monte Carlo simulations, for a non-uniform Zipf probability distribution over the sequences, with error probability $\mathbb{P}_e=0.1$,  for $n=511$, for different values of $k$ and $h$.}{0.45}{t}

%\fig{comp_err_h}{Comparison of the proposed expression from Proposition~\ref{prop:Et_strat3} with Monte-Carlo simulations for a given error probability $P_e=0.1$ that the decoding is incorrect so that the $h$ coupons are erased, for $n=511$, for different values of $k$ and $h$ }{0.45}{t}

%\fig{RS_partial}{Comparison of the proposed expression with Monte-Carlo simulations for partial RS, in the uniform case, with $h=10$, $n=512$, and several values for the number $J$ of sub-blocks.}{0.45}{t}

We first evaluate the accuracy of our theoretical analysis by comparing the expressions provided in Section~\ref{sec:theoretical_expressions} with Monte Carlo simulations. We start with Proposition~\ref{prop:average_T} for the uniform case without errors. We set $h=10$, and consider different values of $n\in \{63,511,8191,65536\}$ in order to evaluate our approximation in different regimes (low $n$ to high $n$). Figure~\ref{fig:comp_ET} provides $E[\widetilde{T}(n,k,h)]$ evaluated either from Proposition~\ref{prop:average_T} (Approx), from the upper bound in~\eqref{eq:upper_bound} (UB), or from Monte Carlo simulations (MC). For all the considered value of $n$, we observe that the upper bound is loose, while the proposed expression coincides with the results of the Monte Carlo simulations. We also measure the following average relative errors between the theoretical expressions and the MC simulations: $0.5\%$ for $n=63$, $0.1\%$ for $n=511$, $0.01\%$ for $n=8191$, $0.003\%$ for $n=65536$. As expected, the relative error decreases as $n$ increases.  This confirms the accuracy of the proposed theoretical expression.

Next, we consider the expression of Proposition~\ref{prop:Et_nonunif} for the non-uniform case without errors. Here, as in~\cite{boneh1997coupon}, we consider the Zipf distribution for the $p_i$, with parameter $s=1.5$.  
Figure~\ref{fig:comp_err_h_nonunif_zipf} compares the theoretical expressions of  $E[\widetilde{T}_{\mathbf{p}}(n,k,h)]$ from Proposition~\ref{prop:Et_nonunif}, with MC simulations, for $n=511$ and $h \in \{4,6,8,10\}$. Again, we observe that the theoretical expression of the proposition coincides with the MC simulations. Besides, the average relative errors range between $0.38\%$ for $h=4$ and $0.55\%$ for $h=10$, which confirms the accuracy of the proposed analysis. Interestingly, we see that the expected number of reads is higher by orders of magnitude than in the uniform case. This comes from the fact that with the Zipf distribution, a non-negligible part of the sequences have a very low sampling probability $p_i$. 

In Figure~\ref{fig:comp_err_h_nonunif_Pe}, we consider a non-uniform distribution in the case with errors, setting an error probability $\mathbb{P}_e=0.1$, and we compare the theoretical expression of~\eqref{eq:Et_strat3} with MC simulations. Here again, we observe that the theoretical expressions coincide with Monte Carlo simulations. 
Also, as expected, we see that the values of $E[T_{\mathbf{p}}(n,k,h)]$ are higher than in the case without errors shown in Figure~\ref{fig:comp_err_h_nonunif_zipf}. The average relative error for each of the four curves is about $0.5\%$.  Although we do not show the curves in the paper for conciseness, we also investigated the uniform case with errors (Proposition~\ref{prop:Et_strat3}), and measured a relative error of about $0.1\%$ for the four considered values of $h$. %This is consistent with the proof for the non-uniform case, which relies on a surrogate distribution. 

%Finally, in Figure~\ref{fig:RS_partial} we consider the setup of Proposition~\ref{prop:block_erasing} with partial RS decoding by sub-block. This time, we consider several values for $J \in \{1,2,4,16,64\}$, the number of sub-blocks in the sequence, and we show the expected number of reads with respect to $P_{\text{err}}$, the sub-block error probability. Again, when comparing to Monte-Carlo simulations, we observe that the theoretical expressions are accurate.  We also see that, as expected, increasing the number of sub-blocks strongly reduces the average number of reads.  On the other hand, in practice, one marker needs to be introduced for each sub-block, which will be taken into account when addressing the optimization problem. 

\subsection{Channel model and codes considered in the optimization}\label{sec:considered_codes}

We now describe the channel model, consensus algorithm, inner codes, and outer codes we consider in our optimization. 

\subsubsection{Channel model}
%Current synthesis technologies introduce negligible error rates compared to sequencing. In contrast, sequencing platforms such as Nanopore and Illumina introduce IDS errors. 
%Many existing works either assume that the DNA storage process introduces a maximum number of errors per sequence~\textcolor{red}{[REF]}, or consider a statistical model with independent and identically distributed (i.i.d.) errors~\textcolor{red}{[REF]}. 
We adopt the channel model introduced in~\cite{hamoum2023channel} and considered in~\cite{maarouf2023achievable}, which targets Nanopore sequencing and captures both the statistics and memory of sequencing errors (insertions, deletions, substitutions) through a hidden Markov model. This model was trained on several datasets, yielding different error profiles with error rates ranging from $0.9\%$ to $10\%$. Here, we consider two error profiles with error-rates $1\%$ and $4\%$, respectively. Note that in this work, the error model does not impact the theoretical developments, and we consider Nanopore sequencing as a relevant use-case to provide numerical optimization results.

\subsubsection{Consensus algorithm}
As in~\cite{park2023reducing,hanna2026reliability}, we consider multiple-sequence alignment of the $h$ reads, here implemented with MAFFT~\cite{katoh2008recent}, followed by majority voting to produce the output consensus sequence. We observed in our numerical results that while more complex, this consensus algorithm provides improved performance compared to algorithms using local alignments between pairs of sequences, like OLC~\cite{li2012comparison} and CCSA~\cite{de2025bruijn}. %Global alignment is known to be more complex, but in DNA data storage, the number $h$ of sequences is rather low (less than $20$), and hence complexity is not an issue. 

\subsubsection{Outer codes}
We consider both RS codes and Raptor codes~\cite{shokrollahi2006raptor}, which today represent the  dominant families of outer codes for DNA data storage. RS codes achieve minimum redundancy, but RS erasure decoding is complex. By contrast, Raptor codes have decoding complexity linear with $n$, owing to the underlying sparse graph structure. 
Therefore, we are interested in quantifying the loss in terms of $\mathbb{E}[T_{\mathbf{p}}(n,k,h)]$ for Raptor codes compared to RS codes.  
%Therefore, we consider them in our analysis to quantify the loss in terms of $\mathbb{E}[T_{\mathbf{p}}(n,k,h)]$  compared to RS codes.  %But Raptor codes typically require $k > m$ sequences to reconstruct the full information. 
%\textcolor{blue}{RS codes are Maximum Distance Separable (MDS), and it is shown in~\cite{bar2024cover} that MDS codes minimize the average number of reads over all possible codes. }

\subsubsection{Inner codes}
The choice of ECC is considerably more open, as many codes have been proposed for DNA data storage. To cover a range of options, we consider RS codes, LDPC codes from~\cite{hamoum2023synchronization} (hard decoding) and~\cite{shibata2019design} (soft decoding), convolutional codes~\cite{maarouf2022concatenated}, MGCP codes~\cite{khabbaz2025marker}, and IDS codes~\cite{song2022systematic,sima2020optimal}. RS codes correct substitution errors only, while the other codes can handle insertions, deletions, and substitutions. 
%For LDPC codes, we consider only regular codes with $d_v =3$, and evaluate several different rates corresponding to codes with different values $d_c$. 
%For convolutional codes, we consider the solution of~\cite{maarouf2022concatenated}, which does not use a consensus algorithm. We use two solutions from ~\cite{maarouf2022concatenated}: convolutional codes alone with rate $R=1/2$, and convolutional codes concatenated with LDPC codes, with overall rate $3/8$. 
% 
IDS codes from~\cite{song2022systematic,sima2020optimal} are designed to correct a fixed number of errors, while convolutional, LDPC, and MGCP codes target statistical models. The specific parameters we consider for each code are provided in Table~\ref{tab:ECC_config}. In this table, we also provide the acronyms used in the tables of results, as well as the parameters considered for the non-optimized setups for comparison. %Interestingly, these works provide scaling laws of the redundancy $r = N-L$ with respect to $N$ and with respect to $t$, the number of errors the code can correct. For instance, the coding solution of~\cite{sima2020optimal} can correct $t$ errors with a redundancy that scales as $r = 4t\log(N)$. %, given that the existential lower bound scales as $r=2t\log(N)$ . 

\subsection{Optimization results (Inner codes)}

\begin{table*}[htbp]
\caption{ECC configurations considered in the optimization}
\centering
\label{tab:ECC_config}
\begin{tabular}{l|c|c}
\toprule
Acronym & Configuration & Non-opt setup \\
\midrule
%c+LB & Consensus algorithm~\cite{katoh2008recent}, lower bound $r=2t\log(N)$ on the redundancy of IDS codes & $h=5$, $t=3$ \\
%\midrule
CC+LDPC & Concatenated construction of~\cite{maarouf2022concatenated}, with CC code $(7,5)$ of rate $1/2$ and regular $(3,12)$ LDPC code of rate $3/4$ & $h=3$ \\
\midrule
c+MGCP & Consensus algorithm~\cite{katoh2008recent} and MGCP codes of~\cite{khabbaz2025marker} & $h=3$, $p_c= 4$ \\
\midrule
c+LDPC (hard) & Consensus algorithm~\cite{katoh2008recent} and LDPC codes from~\cite{hamoum2023synchronization}, with regular codes with $d_v=3$  & $h=4$, $(3,12)$ \\
\midrule
c+LDPC (soft) & Consensus algorithm~\cite{katoh2008recent} and LDPC codes without markers from~\cite{shibata2019design}, with degree distributions from~\cite[Table II]{shibata2019design}  & $h=3
$, $R=0.65$ \\
\midrule
CC & CC decoder from~\cite{maarouf2022concatenated} with CC code $(7,5)$ of rate $1/2$ & $h=3$ \\
\midrule
c+IDS & Consensus algorithm~\cite{katoh2008recent}, IDS code from~\cite{sima2020optimal}, with redundancy $r=4t\log(N)$ & $h=5$, $t=3$ \\
\midrule
c+RS & Consensus algorithm~\cite{katoh2008recent}, RS code to only correct substitution errors & $h=5$, $t=3$ \\
\bottomrule
\end{tabular}
\end{table*}

Now, we investigate the optimization problem introduced in Section~\ref{sec:optim_pb}. At first, we consider RS codes for the outer code, and we compare the different inner codes  presented in Section~\ref{sec:considered_codes}. We set parameters $N=512$ (number of bits in each DNA strand) and $n=511$ (number of DNA strands). We further consider different values of $V_b \in \{ 91980,130305,170352,196735\}$, which correspond to different scenarios in terms of available redundancy, from highest redundancy for $V_b=91980$ bits to lowest redundancy for $V_b=196735$ bits.

First, in Table~\ref{tab:results_4p} we present the optimization results for the channel  model from~\cite{hamoum2023channel} with an error rate of about $4\%$ and considering a uniform distribution for the DNA strands. The column $\mathbb{E}[T]^{\star}$ shows the minimum expected number of reads which can be achieved by the corresponding inner code. For each value $V_b$, we ordered the inner codes from the lowest $\mathbb{E}[T]^{\star}$ to the highest $\mathbb{E}[T]^{\star}$. Interestingly, we observe that LDPC codes (soft) from~\cite{shibata2019design}
provide the best solution for $V_b=91980$ and $V_b=130305$, while becoming third when $V_b$ increases to $170352$ and disappearing for $V_b=196735$.  Besides, while the CC+LDPC solution from~\cite{maarouf2022concatenated} is second for $V_b=91980$, it then disappears from the next values of $V_b$, as it does not satisfy the constraint on $V_b$. We observe that LDPC codes (hard) from~\cite{hamoum2023synchronization}, IDS codes from~\cite{sima2020optimal}, and RS codes for correcting substitutions are the only ones that can provide a valid solution for any value of $V_b$. 
In Table~\ref{tab:results_4p}, we also provide a comparison with the non-optimized setups for each ECC, given by the column $\mathbb{E}[T]^{\text{no}}_{\text{opt}}$. We see that the gains at optimizing each solution are significant in terms of average number of reads. 

In Table~\ref{tab:results_1p}, we provide optimization results for the uniform case with the channel model with $1\%$ error rate. For the first two values of $V_b$, we observe that three solutions perform very close to each other: LDPC (soft), MGCP, and LDPC (hard). Besides, while the CC+LDPC solution was second for $V_b=91980$ when considering the $4\%$ error model, it is now last. This comes from the fact that this solution requires a large amount of redundancy for the ECC, which is not useful anymore when the error probability is as low as $1\%$.  For $V_b=170352$, we see that while LDPC (hard) and MGCP are still competitive, LDPC (soft) ranks last, most probably also due to its increased redundancy for the ECC. We also observe that in many cases, the best value of $h$ is $h^{\star}=1$, which means that the consensus algorithm is not useful when considering a $1\%$ error rate. 

In Table~\ref{tab:results_4p_nf}, we consider the non-uniform case with the channel model at $4\%$ error rate, and we consider a Zipf distribution with $s=1.5$ for $\mathbf{p}$. This time, we observe that LDPC (soft) significantly outperforms the other codes for the first two values of $V_b$, while again falling short for $V_b=170352$. Also, again, LDPC (hard) and MGCP are close to each other for the first two values of $V_b$. 
In this case, like for the other ones, we observe that the optimal values for $h$ tend to increase with $V_b$. 
For small values of $V_b$, given the cost of collecting multiple reads for each unique sequence, it is more interesting to let the inner code handle errors given that the redundancy budget is high. 
However, as $V_{b}$ increases, the redundancy space available for the erasure code and the ECC shrinks, which leads to larger value for $h$. 

The diversity of optimization results, depending on the error-rate, the value of $V_b$, the uniform/non-uniform distribution of the molecules, illustrates the need to select an optimal combination of codes and parameters for the considered operational conditions.  
We also emphasize that these optimization results only compare codes from the perspective of the read cost. Especially, the best codes in terms of average number of reads, are not necessarily the best ones in terms of error probability. For example, we observed from our results that the CC+LDPC solution from~\cite{maarouf2022concatenated} has the best frame-error-rate over all considered codes, but it is not necessarily a good solution in terms of coverage depth, due to its low rate $R=3/8$.

\begin{table}[htbp]
\centering
\caption{Comparison of different inner codes for $N=512, n=511$, channel model with $4\%$ error rate. We consider RS codes, and the uniform distribution. Solutions that do not match the constraint on $V_{\text{b}}$ are not included in the table. The last column provides $\mathbb{E}[T]$ in the non-optimized setup. }
\label{tab:results_4p}
%\resizebox{1\columnwidth}{!}
{\begin{tabular}{llccccc}
\toprule
$V_{\text{b}}$ & \textbf{ECC} & \textbf{$h^{\star}$} & \textbf{$k^\star$} & \textbf{$L^{\star}$} & \textbf{$\mathbb{E}[T]^{\star}$} & \textbf{$\mathbb{E}[T]^{\text{no}}_{\text{opt}}$}  \\
\midrule
\multirow{3}{*}{ $91980$}
&  c+LDPC (soft)   & 1 &  274 &   336 & 740 & 1462 \\
%& c+LB & 3 & 238 & 387 & 1392 & 2167 \\
& CC+LDPC   & 1 & 480 & 192 & 1484 & 3045 \\
& c+MGCP & 3 & 250 & 368 & 1547 & 1671\\
& c+LDPC (hard)   & 3 & 257 & 358 & 1723 & 3118 \\ 
& CC   & 2 & 360 & 256 & 1726 & 1977 \\
& c+IDS & 3 & 295 & 312 & 1854 & 2257 \\
%&  Raptor & c+LDPC   & 3 & 315 & 292 & 2348 \\ 
& c+RS     & 6 & 183 & 503 & 3003 & 3314 \\
 
\midrule

\multirow{3}{*}{ $130305$}
%& c+LB   & 3 & 337 & 387 & 1896 & 2638 \\
&  c+LDPC (soft)   & 1 & 388 & 336   & 1369 & 2068 \\
& c+MGCP   & 3 & 355 & 368 & 2245 & 2437 \\
& c+LDPC (hard)   & 3 & 364 & 358 & 2525 & 4079 \\  
& c+IDS   & 4 & 337 & 387 & 2594 & 2795 \\
%& Raptor & c+LDPC   & 4 & 340 & 384 & 3633 \\ 
& c+RS     & 6 & 259 & 505 & 3756 & 4598  \\
& CC   & 3 & 510 & 256 & 5465 & 5465 \\ 
 
\midrule

\multirow{3}{*}{ $170352$}
& c+IDS   & 5 & 390 & 437 & 3524 & 3588 \\
& c+LDPC (hard)   & 4 & 418 & 408 & 3624 & 4068 \\ 
&  c+LDPC (soft)   & 1 & 507  &  336 & 4311 & 4389 \\
& c+MGCP   & 3 & 474 & 360 & 4440 & 4440 \\
%& Raptor & c+LDPC   & 4 & 340 & 384 & 3633 \\ 
& c+RS     & 6 & 336 & 507 & 4858 & 4871  \\
 
\midrule

\multirow{2}{*}{ $196735$}
%& c+LB   & 4 & 451 & 437 & 3647 & 3668 \\ 
& c+IDS   & 5 & 451 & 437 & 4456 & 4456 \\ 
& c+LDPC (hard)   & 5 & 483 & 408 & 5009 & N/A \\ 
& c+RS    & 8 & 387 & 509 & 5758 & 8184  \\
%\midrule
%\multirow{2}{*}{ $208320$}
%& c+LB   & 5 & 451 & 462 & 4130 & 4241 \\ 
%& c+UB   & 6 & 451 & 462 & 4977 & 5185 \\
%& c+RS     & 8 & 411 & 507 & 6209 & 9211  \\
%& c+LDPC (hard)   & 4 & 480 & 434 & 7292 & N/A \\  
%& Raptor & LDPC  & X & X & X & X \\
\bottomrule
\end{tabular}}
\end{table}

\begin{table}[htbp]
\centering
\caption{Comparison of different inner codes for $N=512, n=511$, channel model with $1\%$ error rate. We consider RS codes, and the uniform distribution. Solutions that do not match the constraint on $V_{\text{b}}$ are not included in the table. }
\label{tab:results_1p}
%\resizebox{1\columnwidth}{!}
{\begin{tabular}{llcccc}
\toprule
$V_{\text{b}}$ & \textbf{ECC} & \textbf{$h^{\star}$} & \textbf{$k^\star$} & \textbf{$L^{\star}$} & \textbf{$\mathbb{E}[T]^{\star}$}   \\
\midrule
\multirow{3}{*}{ $91980$}
%& c+LB & 1 & 246 & 374 & 403  \\
& c+LDPC (soft)   & 1 & 274 &  336  & 444  \\
& c+MGCP & 1 & 245 & 376 & 450 \\
& c+LDPC (hard)   & 1 & 240 & 384 & 484  \\
&c+IDS & 1 & 273 & 337 & 633 \\
& CC   & 1 & 360 & 256 & 907  \\
& c+RS  & 3 & 182 & 507 & 1256 \\
& CC+LDPC   & 1 & 480 & 192 & 1406 \\
%&  Raptor & c+LDPC   & 3 & 315 & 292 & 2348 \\ 
 
\midrule

\multirow{3}{*}{ $130305$}
%& c+LB   & 1 & 337 & 387 & 696 \\
& c+LDPC (soft)   & 1 & 388  & 336 & 807  \\ 
& c+MGCP   & 1 & 347 & 376 & 831  \\
& c+LDPC (hard)   & 1 & 340 & 384 & 836  \\ 
& c+IDS   & 1 & 360 & 362 & 1169 \\
%& Raptor & c+LDPC   & 4 & 340 & 384 & 3633 \\ 
& c+RS & 3 & 258 & 507 & 1636  \\
& CC   & 2 & 510 & 256 & 3770 \\ 
 
\midrule
\multirow{3}{*}{ $170352$}
& c+LDPC (hard)   & 1 & 384 & 444 & 1547  \\ 
& c+MGCP   & 1 & 454 & 376 & 1933  \\
& c+IDS   & 3 & 350 & 487 & 1962  \\
& c+RS     & 3 & 336 & 507 & 2126   \\
& c+LDPC (soft)   & 1 & 507   & 336 & 2545  \\ 
 
\midrule

\multirow{2}{*}{ $196735$}
%&  c+LB   & 1 & 464 & 424 & 1966  \\ 
&c+IDS   & 3 & 404 & 487 & 2370 \\ 
&c+RS    & 3 & 389 & 507 & 2589   \\
&  c+LDPC (hard)   & 3 & 454 & 434 & 3056  \\   
%\midrule
%\multirow{2}{*}{ $208320$}
%& c+LB   & 3 & 428 & 487 & 2427 \\ 
%& c+UB   & 3 & 428 & 487 & 2615 \\
%& c+RS     & 3 & 411 & 507 & 2841   \\
%& c+LDPC (hard)   & 3 & 480 & 434 & 3650  \\  
%& Raptor & LDPC  & X & X & X & X \\
\bottomrule
\end{tabular}}
\end{table}

\begin{table}[htbp]
\centering
\caption{Comparison of different inner codes for $N=512, n=511$, channel model with $4\%$ error rate. We consider RS codes, and the non-uniform Zipf distribution. Solutions that do not match the constraint on $V_{\text{b}}$ are not included in the table. }
\label{tab:results_4p_nf}
%\resizebox{1\columnwidth}{!}
{\begin{tabular}{llcccc}
\toprule
$V_{\text{b}}$ & \textbf{ECC} & \textbf{$h^{\star}$} & \textbf{$k^\star$} & \textbf{$L^{\star}$} & \textbf{$\mathbb{E}[T]^{\star}$}   \\
\midrule
\multirow{3}{*}{ $91980$}
%& c+LB & 3 & 231 & 399 & 22411  \\
&  c+LDPC (soft)   & 1 & 274  & 336 & 11515 \\
& c+MGCP & 3 & 250 & 368 & 28490\\
& c+LDPC (hard)   & 3 & 240 & 384 & 28913 \\ 
&  c+IDS & 4 & 224 & 412 & 32521  \\
&  CC   & 2 & 360 & 256 & 37435  \\
& c+RS     & 6 & 182 & 507 & 41681 \\
& CC+LDPC   & 1 & 480 & 192 & 42501 \\
 
\midrule

\multirow{3}{*}{ $130305$}
& c+LDPC (soft)   & 1 &  388  & 336 &  29170 \\
%& c+LB   & 3 & 327 & 399 & 41122 \\
& c+LDPC (hard)   & 3 & 340 & 384 & 55007 \\
& c+MGCP   & 3 & 355 & 368 & 55425 \\
& c+IDS   & 4 & 317 & 412 & 57892\\
%& Raptor & c+LDPC   & 4 & 340 & 384 & 3633 \\ 
& c+RS     & 6 & 258 & 507 & 71088 \\
& CC   & 3 & 510 & 256 & 206839  \\ 
 
\midrule

\multirow{3}{*}{ $170352$}
& c+IDS   & 5 & 369 & 462 & 94053  \\
& c+LDPC (hard)   & 4 & 418 & 408 & 98636  \\ 
& c+RS     & 6 & 336 & 507 & 110609   \\
& c+MGCP   & 3 & 474 & 360 & 139065  \\
&  c+LDPC (soft)   & 1 &  507  & 336  &  147048 \\

 \midrule

\multirow{2}{*}{ $196735$}
%& c+LB   & 4 & 439 & 449 & 108028  \\ 
&c+IDS   & 5 & 426 & 462 & 131243 \\ 
& c+RS    & 6 & 389 & 507 & 148886  \\
& c+LDPC (hard)   & 3 & 454 & 434 & 170446 \\ 
%\midrule
%\multirow{2}{*}{ $208320$}
%& c+LB   & 5 & 440 & 474 & 127796  \\ 
%& c+UB   & 6 & 428 & 487 & 152923 \\
%& c+RS  & 6 & 411 & 507 & 170540   \\
%& c+LDPC (hard)   & 4 & 480 & 434 & 239133 \\  
%& Raptor & LDPC  & X & X & X & X \\
\bottomrule
\end{tabular}}
\end{table}

\subsection{Optimization results (outer codes)}
Now, we want to compare RS codes and Raptor codes for the outer code. For this evaluation, we only consider LDPC codes (hard) for the inner code, and we set the same parameters as before.  In Table~\ref{tab:results_4p_ER}, we provide the optimal $\mathbb{E}[T]$ for all the considered strategies, for two values $V_b = 91980$ and $V_b = 130305$. 
%We observe that combining RS codes with LDPC codes consistently provides the lowest $\mathbb{E}[T]$ for any considered value of $V_{\text{b}}$. 
As expected, we see that replacing RS codes with Raptor codes increases the expected number of reads. Besides, due to their increased redundancy, no valid Raptor configuration is available for the larger values  $V_b = 170352$ and $V_b=196735$. However, Raptor codes can  be a relevant choice in cases where the decoding complexity matters and/or whenever there is a large space left for the redundancy. 

% We intentionally did not include Raptor codes for the largest values of $V_{\text{bits}}$, since for these values, the Raptor code becomes useless in that $k=n$ reads are required to recover the full information.

\begin{table}[htbp]
\centering
\caption{Comparison of outer codes for $N=511, n=511$, channel model with $4\%$ error rate, uniform distribution. We consider LDPC codes (hard) as inner codes here.  Solutions that do not match the constraint on $V_{\text{bits}}$ are not included in the table. }
\label{tab:results_4p_ER}
%\resizebox{1\columnwidth}{!}
{\begin{tabular}{llcccc}
\toprule
$V_{\text{b}}$ & \textbf{ER} & \textbf{$h^{\star}$} & \textbf{$k^\star$} & \textbf{$L^{\star}$} & \textbf{$\mathbb{E}[T]^{\star}$} \\
\midrule
\multirow{2}{*}{ $91980$}
%&  RS part + LDPC (hard)  & 1 & 291 & 317 %& 429 \\ % J=2 
%& RS + LDPC (soft)   & 1 & 426 & 216 & 740  \\
%&  RS part + LDPC (soft)  & 1 & 426 & 248 & 896 \\
& RS + LDPC (hard)  & 3 & 257 & 358 & 1723 \\ 
&  Raptor + LDPC (hard) & 3 & 315 & 292 & 2348 \\ 
 
\midrule

\multirow{2}{*}{ $130305$}
%& RS part + LDPC (hard)   & 1 & 360 & 362 & 619 \\ 
%& RS part + LDPC (soft)   & 1 & 441 & 336 & 1005  \\
%& RS + LDPC (soft)   & 1 & 418 & 312 & 1369  \\
& RS + LDPC (hard)  & 3 & 364 & 358 & 2525 \\  
& Raptor + LDPC (hard) & 4 & 340 & 384 & 3633 \\ 
 
%\midrule

%\multirow{2}{*}{ $170352$}
%&  RS part + LDPC (hard) & 1 & 443 & 336 & 1019 \\ 
%& RS +LDPC (hard)  & 4 & 418 & 408 & 3624  \\ 
%& RS + LDPC (soft)  & 1 & 507 & 336 & 4311  \\

%\midrule

%\multirow{1}{*}{ $196735$}
%& RS part + LDPC (hard)  & 1 & 483 & 408 & 1454 \\ 
%& RS + LDPC (hard) & 4 & 454 & 434 & 5009 \\ 
  
%\midrule
%\multirow{1}{*}{ $208320$}
%& RS part   & 1 & 511 & 408 & 2840 \\ 
%& RS   & 4 & 480 & 434 & 7292 \\ 
\bottomrule
\end{tabular}}
\end{table}

\section{Conclusion}
In this paper, we addressed the minimization of the read cost of a DNA storage system by optimizing the tradeoff between the coding rates of the inner and outer codes. Using Taylor expansions, we derived analytical approximations of the average number of reads required to recover the data, under both uniform and non-uniform sampling, and both with and without residual errors after inner decoding. Our numerical results show that these approximations accurately predict the average number of reads obtained from Monte Carlo simulations. We also provided optimization results for various pairs of inner and outer codes. For a given pair, optimizing the rate tradeoff significantly reduces the average number of reads. Moreover, the most suitable inner code depends on the operating regime, that is, on the available redundancy budget and the channel error rate. 
Future work will include investigating the synthesis cost, and developing an optimization method from the full distribution of the average number of reads, rather than just the expectation.

%In this paper, we addressed the optimization of the read cost of a DNA storage system, through the tradeoff between  the coding rates of the inner code and the outer code. From Taylor expansions, we provided analytical expressions of the average number of reads needed to reconstruct the data, for both the uniform and the non-uniform case, and in setups with and without errors after ECC decoding. Our numerical results show that the proposed analytical expressions accurately predict the average number of reads measured from Monte-Carlo simulations. We also provided optimization results for various pairs of inner codes and outer codes. These results show that for a given pair, optimizing the tradeoff significantly reduces the average number of reads. We also observed that the inner code needs to be selected depending on the considered regime in terms of available redundancy space and channel error rate.  

\appendices

\section{Proof of Proposition~\ref{prop:average_T} }\label{app:prop1}
The proof builds on~\cite[Theorem~3.1]{holst1986birthday}, which embeds the coupon collector problem in a Poisson process. It considers $n$ Poisson process $\mathcal{P}_j$, each with intensity $1/n$, and interarrival times $W^{(j)}_{1},W^{(j)}_{2}\cdots$, where $W^{(i)}_{i} \sim \mathcal{E}(1/n)$, the exponential distribution with rate $1/n$. The collection is complete as soon as $h$ events have occurred in $k$ of the processes. We use $T_j$ to denote the time until process $\mathcal{P}_j$ has observed $h$ events. The random variables in the sequence $(T_j)_{j \in \llbracket 1,n\rrbracket}$ are i.i.d., and
\begin{equation}
T_j = \sum_{i=1}^h W^{(j)}_i . 
\end{equation}
Using the moment generating function, one can show that $T_j \sim \Gamma(h,n)$. 
Then,~\cite[Theorem~3.1]{holst1986birthday} states that
%which introduces a sequence $(T_j)_{j \in \llbracket 1,n\rrbracket}$ of $n$ i.i.d.\ random variables with $T_j \sim \Gamma(h,1)$,  
\begin{equation}\label{eq:exp_sum}
\mathbb{E}[\widetilde{T}(n,k,h)] = \, \mathbb{E}[T_{k:n}],
\end{equation}
where $T_{k:n}$ denotes the $k$-th order statistic of $(T_j)_{j \in \llbracket 1,n\rrbracket}$. Unfortunately,~\cite{holst1986birthday} does not provide an explicit expression for $\mathbb{E}[T_{k:n}]$.  However, such an expression can be obtained from the theory of order statistics~\cite{arnold2008first}. Consider a sequence of $n$ i.i.d. random variables $(U_j)_{j \in \llbracket 1,n\rrbracket}$, where each $U_j$ follows a uniform distribution $\mathcal{U}([0,1])$, and the $k$-order statistic of the sequence is denoted $U_{k:n}$. Then,~\cite[Eq. (2.4.2)]{arnold2008first} provides that $T_{k:n} \overset{d}{=} \mathbb{F}^{-1}(U_{k:n})$, where $\mathbb{F}$ is the cumulative distribution function of $T_{j}$. Further applying a Taylor expansion on $\mathbb{E}[\mathbb{F}^{-1}(U_{k:n})]$ around $\mathbb{E}[U_{k:n}] = \frac{k}{n+1}$ yields~\cite[Section~5.5]{arnold2008first}
\begin{equation}\label{eq:taylor_exp}
\mathbb{E}[T_{k:n}] = \mathbb{F}^{-1}\!\left(\frac{k}{n+1}\right) + O\!\left(1\right).
\end{equation}
% with~\eqref{eq:exp_sum}, completes the proof. 

\section{Proof of Proposition~\ref{prop:Et_nonunif}}\label{app:prop2}
\subsubsection{Models}
We reuse the notation of the proof of Proposition~\ref{prop:average_T}, and consider two sampling models. The first model is the one we target in the proposition: the $T_j$ are independent but not identically distributed, and $T_j \sim \Gamma(h,1/p_j)$. The number of reads needed to collect $h$ copies of $k$ molecules is given by $T_{k:n}$, the $k$-th order statistic of the $(T_j)_{j\in \llbracket 1,n\rrbracket}$. Unfortunately, in this case, we cannot apply the results of~\cite{arnold2008first} to express $\mathbb{E}[T_{k:n}]$ from a Taylor expansion, because the $T_j$ are not identically distributed. Therefore, we also introduce a surrogate model in which we assume a sequence of $n$ i.i.d. random variables $\overline{T}_j$ such that $\overline{T}_j \sim \mathbb{F}_{\mathbf{p}}$ given in~\eqref{eq:distrib_nonunif}. In this case, since the $\overline{T}_j$ are i.i.d., as in the proof of  Proposition~\ref{prop:average_T}, one can show that the expectation of the $k$-order statistic of the sequence $(\overline{T}_j)_{j\in \llbracket 1,n\rrbracket}$ can be expressed as
\begin{equation}
\mathbb{E}[\overline{T}_{k:n}] = \mathbb{F}_{\mathbf{p}}^{-1}\!\left(\frac{k}{n+1}\right) + O(1).
\end{equation}
We aim to bound the gap between $\mathbb{E}[\overline{T}_{k:n}]$ and $\mathbb{E}[T_{k:n}]$. The proof relies on the two assumptions of the proposition: (i)  there exists $c_1,c_2$ such that $c_1/n\le p_j\le c_2/n$, (ii) there exists $\varepsilon$ such that $k/n\in[\varepsilon,1-\varepsilon]$. 

\subsubsection{Number of completed coupons} 
We are interested in the quantity $Z(x)$ defined as 
\begin{equation}
Z(x) = \sum_{j=1}^n \mathbb{1}(T_j \leq x),
\end{equation}
which represents the number of sequences for which $h$ copies have been collected by time $x$. The event $(Z(x) \leq k-1)$ is equivalent to  $(T_{k:n} > x)$. We also define $\overline{Z}(x) = \sum_{j=1}^n \mathbb{1}(\overline{T}_j \leq x) $.  
From the definition of $Z(x)$ and $\overline{Z}(x)$, one can show that
\begin{align}
\mu(x) = \mathbb{E}[Z(x)] & = \mathbb{E}[\overline{Z}(x)] = \sum_{j=1}^n \mathbb{F}_j(x), \\ 
\sigma^2(x) = \mathbb{V}(Z(x)) &= \sum_{j=1}^n \mathbb{F}_j(x)(1-\mathbb{F}_j(x)), \\ 
\bar{\sigma}^2(x) = \mathbb{V}(\overline{Z}(x)) & = \sum_{j=1}^n \mathbb{F}_j(x) - \frac{1}{n} \sum_{j=1}^n \sum_{j'=1}^n \mathbb{F}_j(x)\mathbb{F}_{j'}(x) , 
\end{align} 
where $\mathbb{V}(.)$ is the variance. The function $\mu:\mathbb{R} \rightarrow \mathbb{R}$ is increasing. For convenience in what follows, we denote $ \overline{\mathbb{F}} := \mathbb{F}_{\mathbf{p}}$.  In addition, $\overline{\sigma}^2 = n\overline{\mathbb{F}}(1-\overline{\mathbb{F}})$, and
$$ \overline{\sigma}^2(x) - \sigma^2(x) = \sum_{j} \mathbb{F}_j^2 - n\mathbb{\overline{F}}^2 = \sum_{j} (\mathbb{F}_j - \overline{\mathbb{F}})^2 \geq 0.$$ 
Hence, $\sigma(x)^2 \leq  \overline{\sigma}^2(x) \leq \frac{n}{4}$. 

\subsubsection{Integral expression}
We express
\begin{equation}\notag
\mathbb{E}[T_{k:n}] = \int_{0}^{+\infty} \mathbb{P}(T_{k:n}\geq x) dx = \int_{0}^{+\infty} \mathbb{P}(Z(x) \leq k-1) dx ,
\end{equation}
and hence
\begin{align}\label{eq:ub_diff}
& \left|\mathbb{E}[T_{k:n}] - \mathbb{E}[\overline{T}_{k:n}]\right|  \\ \notag
& = \left| \int_{0}^{+\infty} \left(\mathbb{P}(Z(x) \leq k-1) - \mathbb{P}(\overline{Z}(x) \leq k-1) \right)dx \right| .
\end{align}
To bound this integral, we define 
\begin{equation}
\mathcal{C}_{\lambda} = \left\{ x > 0, |k-1-\mu(x)| \leq \lambda \overline{\sigma}(x) \right\},
\end{equation}
where $\lambda = \sqrt{3\log n}$. The interval $[x_{-}, x_{+}]$ is defined as the convex hull of $\mathcal{C}_{\lambda}$, and we can show that $x_{-},x_{+} \in \mathcal{C}_{\lambda}$. 
We then express
\begin{align}
& \left|\mathbb{E}[T_{k:n}] - \mathbb{E}[\overline{T}_{k:n}] \right|  \leq \\ \notag
%& \leq A_{]0,x_{-}]} + A_{[x_{-}, x_{+}]} + A_{[x_{+}, Kn]} +  A_{[Kn, +\infty[}, 
&A_{]0,x_1]}+A_{[x_1,x_-]}+A_{[x_-,x_+]}+A_{[x_+,x_2]}+A_{[x_2,+\infty[},
\end{align}
where $A_I$ is the absolute value of the integral in~\eqref{eq:ub_diff} on the interval $I$. We set $x_1$, $x_2$  such that $\mu(x_1)=k/e^2$, $\mu(x_2)=n-(n-k+1)/e^2$, and for $n$ large enough and $\lambda = o(\sqrt{n})$, one can show that $x_1<x_-< x_+<x_2$. In what is next, we investigate the convergence of each integral term separately, through six technical lemmas, as summarized in Table~\ref{tab:summary_proof}. This allows us to conclude that 
\begin{equation}
\left|\mathbb{E}[T_{k:n}] - \mathbb{E}[\overline{T}_{k:n}] \right|  = O(\sqrt{n\log n}).
\end{equation}

\begin{table}[h]
\centering
\begin{tabular}{|c|c|c|c|}
\hline 
 Interval & Bound & Order & Lemma \\
\hline 
$ A_{]0,x_1]} $ & Chernoff & $o(1)$ & 3 \\
$ A_{[x_1,x_{-}]} $ & Bernstein & $O(n^{-1/8})$ & 4 \\
$ A_{[x_{-},x_{+}]} $ & Berry-Esseen & $O(\sqrt{n\log n})$ & 6 \\
$ A_{[x_{+},x_2]} $ & Bernstein & $O(n^{-1/8})$ & 5 \\
$ A_{[x_2,+\infty[} $ & Chernoff & $o(1)$ & 2\\
\hline
\end{tabular}
\caption{Summary of the proof of Proposition~\ref{prop:Et_nonunif}}\label{tab:summary_proof}
\end{table}

\subsubsection{Technical lemmas}
In each of the five regions, we bound $\mathbb P(Z(x)\le k-1)$ and $\mathbb P(\overline Z(x)\le k-1)$ or their complements, according to
\begin{align}
& \mathbb{P}(Z(x) \leq k-1) - \mathbb{P}(\overline{Z}(x) \leq k-1) \\ \notag
& = \mathbb{P}(\overline{Z}(x) >k-1) - \mathbb{P}(Z(x) > k-1) . 
\end{align}
We denote $m=n-k+1$, $\nu(x)=n-\mu(x)$, and $a(x) = k - 1-\mu(x) $. Since $\mu$ is increasing, $a$ is decreasing in $x$. We use $x^{\star}$ to denote the solution of $\mu(x^{\star}) = k-1$. We first bound $x^{\star}$. 

\begin{lemma}
$ x^{\star} = O(n)$ .
\end{lemma}
\begin{proof}
First, $\mathbb{F}_j(x) = G_h(p_jx)$, where $G_h$ is the cdf of the Gamma distribution $\Gamma(h,1)$ and $G_h$ is increasing with $x$. Hence $G_h(c_1x/n)\le\mathbb F_j(x)\le G_h(c_2x/n)$ and
\begin{equation}
nG_h\left(\frac{c_1x}{n} \right) \leq \mu(x) \leq nG_h\left(\frac{c_2x}{n} \right).
\end{equation}
Considering $\mu(x^{\star}) = k-1$ gives 
$$G_h\!\Big(\frac{c_1x^\star}{n}\Big)\le\frac{k-1}{n}\le G_h\!\Big(\frac{c_2x^\star}{n}\Big),$$
and since $G_h^{-1}$ exists and is increasing, 
\begin{equation}
\frac{n}{c_2}G_h^{-1}\left(\frac{(k-1)}{n} \right) \leq x^{\star} \leq \frac{n}{c_1}G_h^{-1}\left(\frac{(k-1)}{n} \right),
\end{equation}
using the condition $k/n \leq 1 - \varepsilon$ and the fact that $G_h^{-1}$ is increasing gives $G_h^{-1}\left(\frac{(k-1)}{n} \right) \leq G_h^{-1}\left(1 - \varepsilon\right)$, 
and hence $x^{\star} = O(n) $. 
\end{proof}

\begin{lemma}
$A_{[x_2,+\infty[} = o(1)$ .
\end{lemma}
\begin{proof}
In the interval $[x_2,+\infty[$, since $x_2 > x_{+}$, $a(x) < - \lambda \overline{\sigma}(x)$, and hence $\nu(x) < m - \lambda \overline{\sigma}(x) < m$. We remark that $n-Z(x)=\sum_j\mathbb 1(T_j>x), $ which allows us to apply the Chernoff bound for a sum of independent Bernoulli variables with mean $\nu(x) < m$. This gives
\begin{equation}
   \mathbb{P}(Z(x)\le k-1) =  \mathbb{P}(n-Z(x)\geq m)  \leq \left(\frac{e\nu(x)}{m}\right)^m .
\end{equation}
Then, for $x>x_2$, $\mu(x) > \mu(x_2) = n - m/e^2$ which gives $\nu(x) < m/e^2$. Accordingly, $e\nu(x)/m \leq 1/e$. Hence, $$ \left(\frac{e\nu(x)}{m}\right)^m \leq\frac{e\,\nu(x)}{m}\,e^{-(m-1)}. $$ 
%\begin{equation}
%\int_{x_2}^\infty\nu(x)\,dx\le\int_0^\infty\sum_j\mathbb P(T_j>x)\,dx=\sum_j\mathbb E[T_j]=\sum_j h/p_j\le h n^2/c_1
%\end{equation}
In addition, 
$$\int_{x_2}^\infty\nu(x)\,dx  \leq \int_{0}^\infty\nu(x)\,dx = \sum_{j=1}^n \frac{h}{p_j} \leq \frac{h n^2} {c_1},$$
where the equality comes from $\nu(x) = \sum_j \mathbb{P}(T_j>x)$ so that $\int_{0}^\infty\nu(x)\,dx = \sum_j \mathbb{E}[T_j] $. 
The same bound applies to $\overline{Z}$ since the Chernoff bound only depends on the mean which is $\mu(x)$ for both $Z$ and $\overline{Z}$.  Given that $m  = n-k+1\geq\varepsilon n$, we get
\begin{equation}
A_{[x_2,+\infty[} \leq \frac{e\,h\,n}{c_1\varepsilon}\,e^{-(\varepsilon n-1)} =  o(1). 
\end{equation}
\end{proof}

\begin{lemma}
$ A_{]0,x_1]} = o(1)$. 
\end{lemma}
\begin{proof}
For $]0,x_1]$, we bound $\mathbb{P}(Z(x) \geq k)$. For $x \leq x_1$, $\mu(x) \leq \mu(x_1) = k/e^2 < k$. Hence, applying the Chernoff bound leads to 
$$ \mathbb{P}(Z(x) \geq k) \leq \left(\frac{e\mu(x)}{k}\right)^k\leq e^{-k}. $$ %Hence $A_{]0,x_1]} \leq x_1 e^{-k} \leq (1-\varepsilon n) e^{-n\varepsilon} = o(1)$.
In addition, $A_{]0,x_1]} \leq x_1 e^{-k} \leq x^{\star} e^{-k} \leq x^{\star} e^{-\varepsilon n} $, 
which gives $A_{]0,x_1]} = O(ne^{-n\varepsilon}) = o(1)$. 

\end{proof}

\begin{lemma}\label{lem:lemma4}
$A_{[x_{1},x_{-}]} = O(n^{-1/8})$. 
\end{lemma}
\begin{proof}
In the left region $[x_1,x_{-}]$, $k-1-\mu(x)\ge\lambda\bar\sigma(x)$.  
By applying Bernstein's inequality,
\begin{align}
\mathbb{P}(Z(x) > k-1 ) & \leq \mathbb{P}(Z(x) - \mu(x) \geq \lambda \overline{\sigma}(x)) \\
& \leq \exp\left( -\frac{\lambda^2 \overline{\sigma}(x)^2}{2\sigma(x)^2 + 2\lambda\overline{\sigma}(x)/3} \right) \\
& \leq \exp\left( -\frac{3}{2}\frac{\lambda^2 \overline{\sigma}(x)}{\lambda +3\overline{\sigma}(x) } \right) ,
\end{align}
where the last inequality comes from $\sigma(x) \leq \overline{\sigma}(x)$, and the same inequality applies to $\mathbb{P}(\overline{Z}(x) > k-1)$. 
In addition, one can show that $\overline{\mathbb{F}}(x) \geq \mu(x_1)/n \geq \varepsilon / e^2$ and that $1-\overline{F}(x) \geq \varepsilon$. This gives $\overline{\sigma}^2 \geq n \varepsilon^2/e^2$, and for $\lambda^2 = 3\log n $, $\lambda^2 \leq n \varepsilon^2/e^2 \leq \overline{\sigma}(x)^2$. 
This gives
\begin{equation}
A_{[x_1,x_{-}]} \leq ((x_{-}) - x_1)\exp\left( -\frac{3}{8}\lambda^2  \right) .
\end{equation} 
With $x_{-} \leq x{\star} = O(n)$ and   $\lambda^2 =  3\log n$, we get
%\begin{equation}
%A_{[x_1,x_{-}]} \leq x^{\star}\exp\left(-\frac{3}{8}\lambda^2 \right) .
%\end{equation}
 $A_{[x_1,x_{-}]} = O(n^{-1/8})$ .
\end{proof}

\begin{lemma}
$A_{[x_{+},x_{2}]} = O(n^{-1/8})$. 
\end{lemma}
\begin{proof}
In the interval $[x_{+},x_2]$, applying again Bernstein's inequality on $\mathbb P(Z\le k-1)\le\mathbb P(\mu-Z\ge\lambda\bar\sigma),$ and using the condition $\lambda \leq \overline{\sigma}(x)$  as in Lemma~\ref{lem:lemma4} gives that 
\begin{equation}
A_{[x_{+}, x_2]} \leq (x_2 - x_{+})\exp\left( -\frac{3}{8}\lambda^2  \right)  .
\end{equation}
Next, $\mu(x_2) = n - m/e^2$ and 
\begin{equation}
G_h\left( \frac{c_1x_2}{n} \right) \leq \frac{\mu(x_2)}{n} \leq 1 - \frac{\varepsilon}{e^2}.
\end{equation}
Hence $x_2 \leq \frac{n}{c_2} G_h^{-1}\left( 1 - \frac{\varepsilon}{e^2} \right) = O(n)$, and $A_{[x_{+},x_2]} \leq x_2 \exp(-3\lambda^2/8) = O(n^{-1/8})$.

\end{proof}

\begin{lemma}
$A_{[x_{-},x_{+}]} = O(\sqrt{n\log n})$ .
\end{lemma}
\begin{proof}
In the region $[x_{-}, x_{+}]$, we apply the Berry-Esseen theorem and get 
\begin{align}\notag
 & A_{[x_{-}, x_{+}]} 
  \leq \left| \int_{x_{-}}^{x_{+}} \Phi\left(\frac{k-1-\mu(x)}{\sigma(x)}\right) - \Phi\left(\frac{k-1-\mu(x)}{\overline{\sigma}(x)}\right)dx\right| \\
  & ~~~~+  C_0\int_{x_{-}}^{x_{+}}  \left(\frac{1}{\sigma(x)} + \frac{1}{\overline{\sigma}(x)}\right) dx,
\end{align}
where $C_0$ is a constant and  $\Phi$ is the cdf of the standard normal distribution. 
The second term in the inequality can be further upper bounded as
\begin{equation}
 C_0\int_{x_{-}}^{x_{+}} \left( \frac{1}{\sigma(x)} + \frac{1}{\overline{\sigma}(x)}\right) dx \leq 2C_0 \frac{|x_{+} - x_{-}|}{\min_{[x_{-}, x_{+}]} \sigma(x)} . 
\end{equation}
 
Next, $\exists \gamma \in ]0,1/2[$ such that $\forall x \in [x_{-},x_{+}] $, $\forall j \in \llbracket 1,n\rrbracket$, $\gamma \leq \mathbb{F}_j(x) \leq 1-\gamma$. Hence
\begin{equation}
n\gamma(1-\gamma)\leq \sigma(x)^2 \leq \frac{n}{4}
\end{equation}
and $\sigma^2(x)= \Theta(n)$. Similarly, $\overline{\sigma}^2(x) =\Theta(n)$. 
In addition, $\forall x \in \mathcal{C}_{\lambda}$, $|k-1-\mu(x)| \leq \lambda\frac{\sqrt{n}}{2}$. Let $J = \{ x: |k-1-\mu(x)| \leq \lambda\frac{\sqrt{n}}{2} \}$.  Since $\mu(x)$ is increasing, $J$ is an interval, and hence $[x_{-},x_{+}] \subseteq J$. By the mean-value theorem,
\begin{equation}
\mu(x_{+}) - \mu(x_{-}) = \mu'(\xi)(x_{+}-x_{-}),
\end{equation}
where $\xi \in ]x_{-},x_{+}[$, $\mu'(x) = \sum_{j=1}^n f_j(x)$, and $f_j$ is the pdf of the Gamma distribution $\Gamma(h,1/p_j)$. In addition $\mu(x_{-}) = k-1-\lambda\overline{\sigma}(x_{-})$ and $\mu(x_{+}) = k-1+\lambda\overline{\sigma}(x_{+})$. Hence $\mu(x_{+}) - \mu(x_{-}) = \lambda(\overline{\sigma}(x_{+}) + \overline{\sigma}(x_{-})) \leq \lambda\sqrt{n}$. 
Hence
\begin{equation}
|x_{+}-x_{-}|\leq \frac{\lambda\sqrt{n}}{\mu'(\xi)}. 
\end{equation}
By differentiating $\mathbb{F}_j$, one can show that $f_j(x) = p_j g_h(p_jx)$, where $g_h$ is the pdf of $\Gamma(h,1)$. Hence $\mu'(\xi) \geq \frac{c_1}{n}ng_{\min} = c_1g_{\min}$, where $g_{\min} := \min_j g_h(p_j\xi)$.   %Hence if $p_j = \Theta\left(\frac{\alpha_n}{n}\right)$, then $ |x_{+}-x_{-}|= \Theta\left( \frac{\lambda\sqrt{n}}{\alpha_n}\right).$ 
This leads to
\begin{equation}
 |x_{+} - x_{-}| \leq \frac{\lambda\sqrt{n}}{ g_{\min} c_1}
\end{equation}
and
\begin{equation}
2C_0 \frac{|x_{+} - x_{-}|}{\min_{[x_{-}, x_{+}]} \sigma(x)} \leq 2C_0 \frac{\lambda\sqrt{n}}{ g_{\min} c_1 \sqrt{n\gamma(1-\gamma)}} = O( \lambda) .
\end{equation}
Besides, one can show that
\begin{align}
& \left| \int_{x_{-}}^{x_{+}} \Phi\left(\frac{k-1-\mu(x)}{\sigma(x)}\right) - \Phi\left(\frac{k-1-\mu(x)}{\overline{\sigma}(x)}\right)dx\right|\\
& \leq \int_{x_{-}}^{x_{+}}  \frac{\overline{\sigma}(x) - \sigma(x)}{\sigma(x)}dx .
\end{align}
Then, on $[x_{-},x_{+}]$, $\sigma(x) \geq \sqrt{n\gamma(1-\gamma)}$ and $\overline{\sigma}(x) \leq \frac{\sqrt{n}}{2}$. Next,  $\frac{\overline{\sigma}(x) -\sigma(x) }{\sigma(x)} \leq \frac{\overline{\sigma}(x)}{\sigma(x)} \leq \frac{1}{2\sqrt{\gamma(1-\gamma)}}$ and
\begin{equation}
\int_{x_{-}}^{x_{+}}  \frac{\overline{\sigma}(x) - \sigma(x)}{\sigma(x)}dx \leq \frac{\sqrt{n}\lambda}{c_1 g_{\min}} \frac{1}{2\sqrt{\gamma(1-\gamma)}} = O(\sqrt{n}\lambda).
\end{equation}
With $\lambda^2 = 3 \log n$, we get $A_{[x_{-},x_{+}]} = O(\sqrt{n\log n})$. 

\end{proof}

%Since $Z(x)$ is defined as a sum of random variables, we apply the Berry-Esseen theorem to express
%\begin{equation}\label{eq:gaussian_approx}
%\mathbb{E}[T_{k:n}] = \int_{0}^{+\infty} \Phi\left(\frac{k-\mu}{\sigma}\right) dx + \epsilon,
%\end{equation}
%where $|\epsilon| \leq \int_{0}^{+\infty}\frac{C_0\sum_{j=1}^n \mathbb{F}_j(x)}{\text{Var}(Z(x))^{3/2}} dx$, $C_0$ is a constant, $\mu = E[Z(x)]$, $\sigma= \sqrt{\text{Var}(Z(x))}$, and $\Phi$ is the cdf of the standard normal distribution. 
%In the same way, 
%\begin{equation}
%\mathbb{E}[\overline{T}_{k:n}]  = \int_{0}^{+\infty} \Phi\left(\frac{k-\mu}{\overline{\sigma}}\right) dx + \overline{\epsilon},
%\end{equation}
%where $\overline{\sigma} = \sqrt{\text{Var}(\overline{Z}(x))}$ and the constant $|\overline{\epsilon}| \leq |\epsilon|$ since $\text{Var}(\overline{Z}(x)) \geq \text{Var}(Z(x))$. 
%This permits to express
%\begin{align}
%&\mathbb{E}[T_{k:n}]  = \mathbb{F}_{\mathbf{p}}^{-1}\!\left(\frac{k}{n+1}\right) + O(1)  \\
%& + \int_{0}^{+\infty} \left(\Phi\left(\frac{k-\mu}{\sigma}\right) - \Phi\left(\frac{k-\mu}{\overline{\sigma}}\right) \right)dx  + 2\epsilon,
%\end{align}
%which concludes the proof. 

\section{Proof of Proposition~\ref{prop:Et_strat3}}\label{app:prop3}
The proof extends that of Proposition~\ref{prop:average_T}.
We still consider $n$ Poisson processes $\mathcal{P}_j$, each with intensity $1/n$, and interarrival times $W^{(j)}_{1},W^{(j)}_{2}\cdots$. In the case with errors, $T_j$, the time at which collection $j$ is complete, is obtained  by observing $R_j$ blocks of $h$ reads each, where $R_j$ is a random variable following a Geometric distribution of parameter $(1-\mathbb{P}_e)$. Since the $R_j$ are i.i.d., the sequence $(T_j)_{j \in \llbracket 1,n\rrbracket}$ is still i.i.d., and now, 
\begin{equation}
T_j = \sum_{i=1}^{hR_j} W^{(j)}_i .
\end{equation}
We express the moment generating function of $T_j$ as
\begin{align}
\mathbb{E}[e^{sT}] &= \mathbb{E}\left[\mathbb{E}\left[e^{s\sum_{i=1}^{hR_j} W^{(j)}_i }|R_j\right]\right] \\
& = \mathbb{E}\left[ \mathbb{E}\left[e^{sW}\right]^{hR_j}\right] \\
& = \sum_{r\geq 1} (1-\mathbb{P}_e) \mathbb{P}_e^{r-1} \left( \frac{1}{1-ns} \right)^{hr}, \label{eq:mgf_G}
\end{align}
where $\left( \frac{1}{1-ns} \right)^{hr}$ is the moment generating function of the Gamma distribution $\Gamma\left(rh,n\right)$.
We identify~\eqref{eq:mgf_G} as the moment generating function of the distribution with cdf $\mathbb{G}$ in~\eqref{eq:G_cdf}, which, by uniqueness, allows us to conclude that the cdf of each $T_j$ is $\mathbb{G}$. 
Then, applying the Taylor expansion from~\cite[Section~5.5]{arnold2008first} on $\mathbb{E}[{T_{k:n}}] = \mathbb{E}[\mathbb{G}^{-1}(U_{k:n})]$ concludes the proof. 

\bibliographystyle{IEEEtran}
\bibliography{sample}

\end{document}